\immediate\write18{makeindex Manuscript_EN.nlo -s nomencl.ist -o Manuscript_EN.nls}
\documentclass[conference]{cas-dc} 
\pdfoutput=1
\newcommand{\cmark}{\ding{51}}
\newcommand{\xmark}{\ding{55}}

\usepackage{graphicx}
\usepackage{xcolor}
\usepackage{gensymb}
\usepackage{siunitx}
\usepackage{float}
\usepackage{amsmath}
\usepackage{longtable}
\usepackage{caption}
\usepackage[utf8]{inputenc}
\usepackage[T1]{fontenc}
\usepackage{amssymb}
\usepackage{lscape}
\usepackage[outline]{contour}
\usepackage{multirow}
\usepackage{subcaption}
\usepackage{pifont}
\usepackage[numbers]{natbib}
\usepackage{gensymb}
\usepackage{xpatch}
\usepackage{lipsum}
\usepackage{nomencl}

\makenomenclature

\definecolor{darkmagenta}{rgb}{0.55, 0.0, 0.55}

\begin{document}

\let\WriteBookmarks\relax
\def\floatpagepagefraction{1}
\def\textpagefraction{.001}

\shorttitle{A Survey on Cellular-connected UAVs: Design Challenges, Enabling 5G/B5G Innovations, and Experimental Advancements}
\shortauthors{D. Mishra et~al.}

\title[mode = title]{A Survey on Cellular-connected UAVs: Design Challenges, Enabling 5G/B5G Innovations, and Experimental Advancements}                      

\author{Debashisha Mishra}\cormark[1]  
\cortext[cor1]{Corresponding author.}

\author{Enrico Natalizio}

\address{Université de Lorraine, CNRS, LORIA, France}

\renewcommand{\printorcid}{}
\nonumnote{E-mails: debashish216@gmail.com, enrico.natalizio@loria.fr}

\begin{abstract}[S U M M A R Y]
As an emerging field of aerial robotics, Unmanned Aerial Vehicles (UAVs) have gained significant research interest within the wireless networking research community. As soon as national legislations allow UAVs to fly autonomously, we will see swarms of UAV populating the sky of our smart cities to accomplish different missions: parcel delivery, infrastructure monitoring, event filming, surveillance, tracking, etc. The UAV ecosystem can benefit from existing 5G/B5G cellular networks, which can be exploited in different ways to enhance UAV communications. Because of the inherent characteristics of UAV pertaining to flexible mobility in 3D space, autonomous operation and intelligent placement, these smart devices cater to wide range of wireless applications and use cases. 
This work aims at presenting an in-depth exploration of integration synergies between 5G/B5G cellular systems and UAV technology, where the UAV is integrated as a new aerial User Equipment (UE) to existing cellular networks. In this integration, the UAVs perform the role of flying users within cellular coverage, thus they are termed as cellular-connected UAVs (a.k.a. UAV-UE, drone-UE, 5G-connected drone, or aerial user). The main focus of this work is to present an extensive study of integration challenges along with key 5G/B5G technological innovations and ongoing efforts in design prototyping and field trials corroborating cellular-connected UAVs. This study highlights recent progress updates with respect to 3GPP standardization and emphasizes socio-economic concerns that must be accounted before successful adoption of this promising technology. Various open problems paving the path to future research opportunities are also discussed. 
\end{abstract}


\maketitle

\section{Introduction} \label{intro}

\sloppy

\nomenclature{UAV}{Unmanned Aerial Vehicle}
\nomenclature{UAS}{Unmanned Aerial System}
\nomenclature{CAGR}{Compound Annual Growth Rate}
\nomenclature{IIoT}{Industrial Internet of Things}
\nomenclature{CPS}{Cyber Physical System}
\nomenclature{AI}{Artificial Intelligence}
\nomenclature{AR/VR}{Augmented Reality/Virtual Reality}

Unmanned Aerial Vehicles, abbreviated as UAVs, are aircrafts without any human pilot onboard, mainly controlled and managed remotely or via embedded autonomous computer programs. UAVs are also popularly known as drones. It is a new paradigm emerged from aerial robotics with enormous potential for enabling new applications in diverse areas and business opportunities~\cite{valavanis2015handbook,beard2012small,asadpour2014micro}. The global UAV market was valued at US\$ $20.68$ billion in $2017$ and is expected to reach US\$ $59.82$ billion by $2026$, at a compound annual growth rate (CAGR) of $14.20$\% during a forecast period~\cite{globaluavmarket}. 


Unique features of UAVs pertaining to high mobility in three-dimensional space, autonomous operation, flexible deployment tend to find appealing solutions for wide range of applications including civil, public safety, Industrial IoT platforms (IIoT), security and defence sectors, cyber physical systems, atmospheric and environmental observation etc~\cite{hayat2016survey,shakeri2019design,wang2019survey}. By leveraging other emerging technologies like Artificial Intelligence(AI), Internet of Things (IoT), Augmented Reality/Virtual Reality(AR/VR), UAVs have been able to showcase substantial value proposition to a wide range of civil and industrial applications across diverse areas. The UAVs are flying platforms with adaptive altitude support and hence, the emerging use cases for each of the mentioned applications demand a secure, reliable wireless communication infrastructure for command and control, as well as an efficient information dissemination towards the ground control station~\cite{mozaffari2019tutorial}. 

\begin{figure}[b]
\centering
\includegraphics[width=1\linewidth]{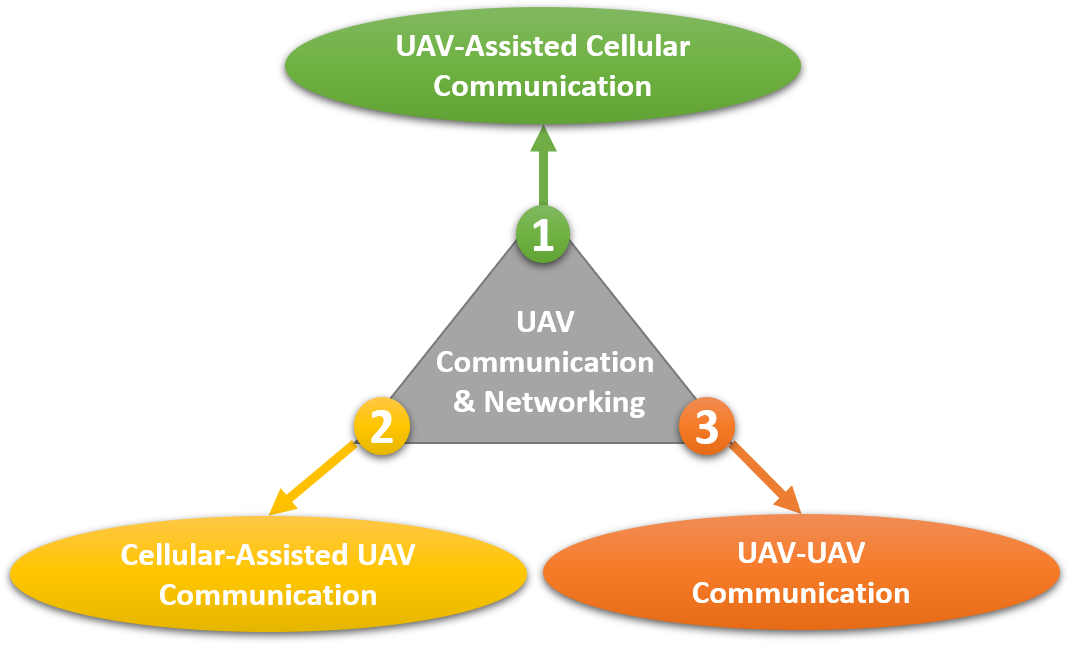}
\caption{Integration opportunities of UAV with Cellular Network}
\label{classification}
\end{figure}

%


\begin{figure*}[htb!]
        \begin{subfigure}[b]{0.32\textwidth}        	   	
                \includegraphics[width=\textwidth]{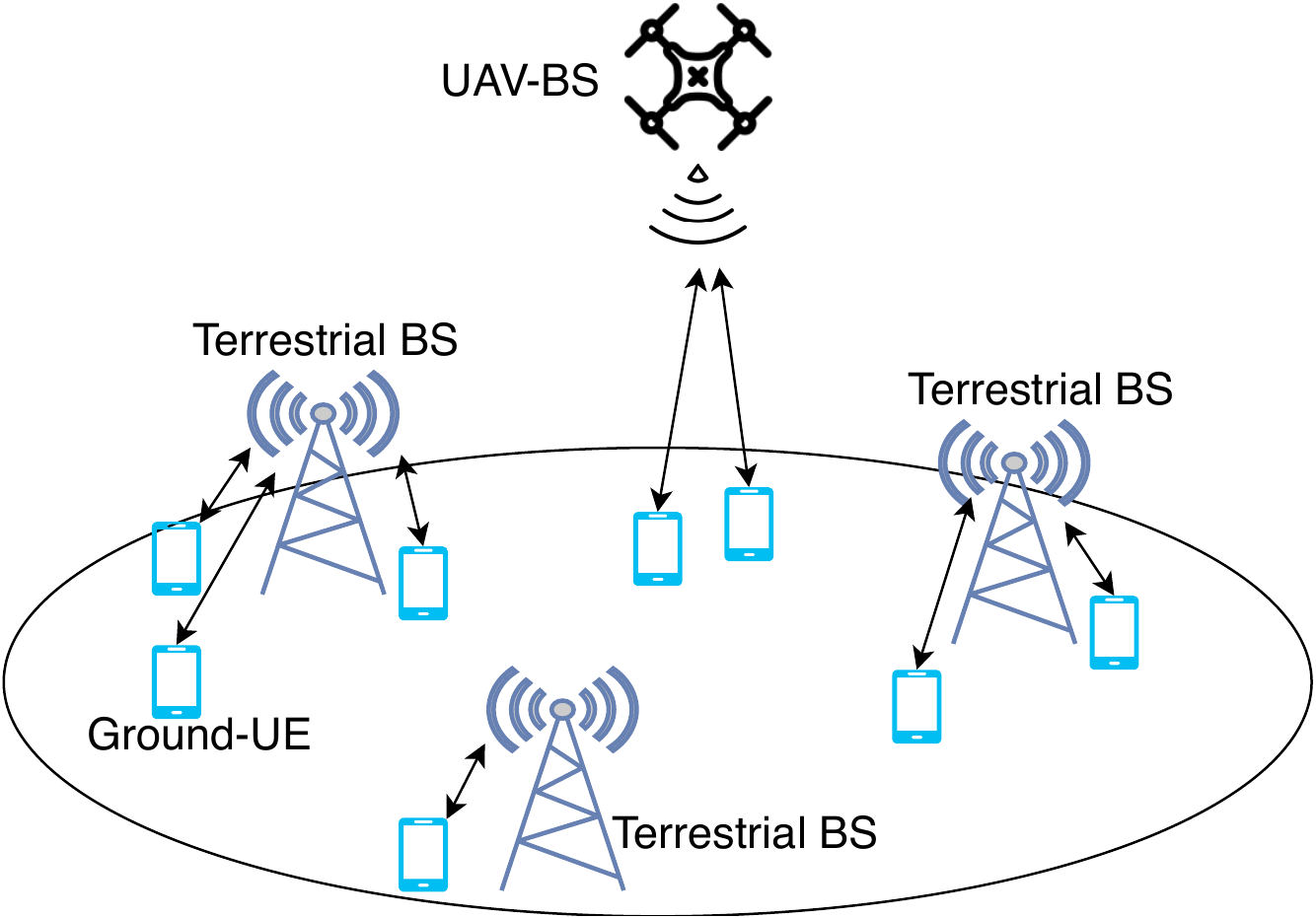}
                \caption{UAV-Assisted Cellular Communication}
                \label{fig:uav-assisted-cellular}
        \end{subfigure} \hfill
        \begin{subfigure}[b]{0.32\textwidth}    
                \includegraphics[width=\textwidth]{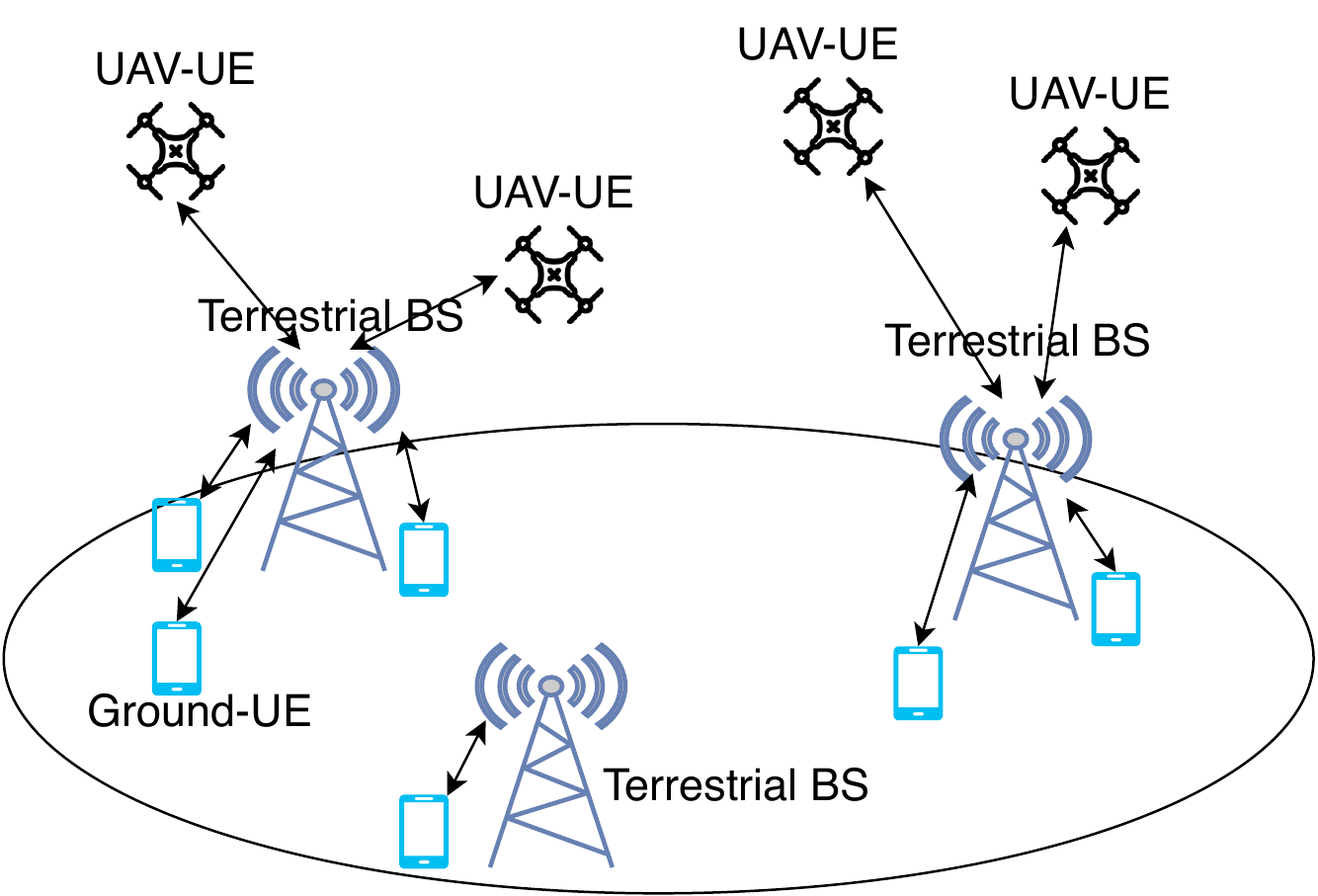}
                \caption{Cellular-Assisted UAV Communication}
                \label{fig:cellular-assisted-uav}
        \end{subfigure}  \hfill
        \begin{subfigure}[b]{0.32\textwidth}
                \includegraphics[width=\textwidth]{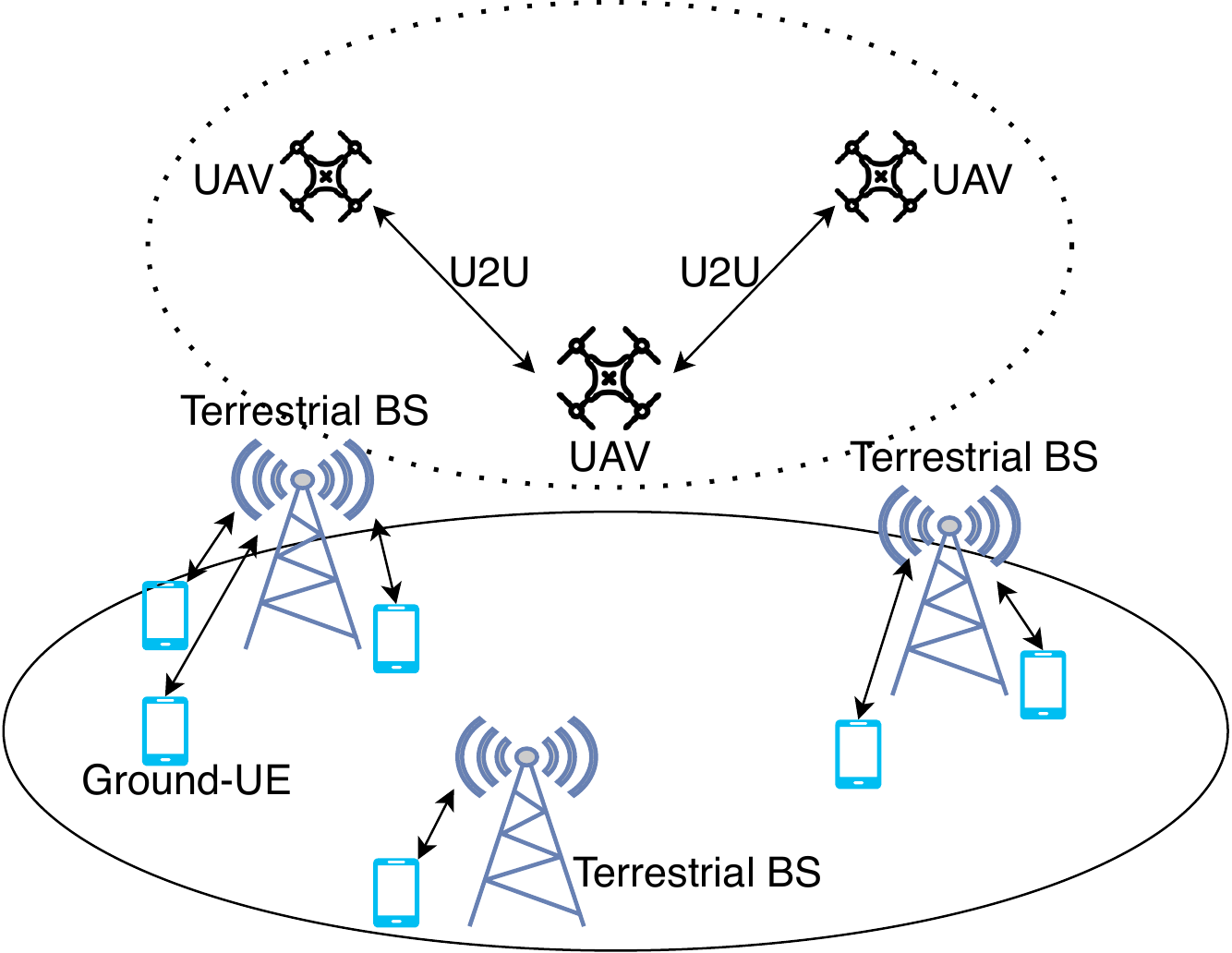}
                \caption{UAV-UAV Communication}
                \label{fig:uav-uav}
        \end{subfigure} \hfill
        \caption{Integration Opportunities of UAV to Cellular Network}\label{fig:intopr}
\end{figure*}


On this advent, there are two main research directions to be investigated. First, how to integrate a suitable wireless communication platform into UAVs for ubiquitous connectivity and seamless service for the identified use cases. Second, what are the scientific and technological challenges that arise from such integration. We aim to focus on both the directions in this paper and highlight several distinctive characteristics, challenges with state-of-the-art solutions from the viewpoint of aerial networking. 

UAVs are inherently mobile in nature and hence, require wireless support for communication needs~\cite{zeng2016wireless}. The wireless communication infrastructure can be provided over licensed or unlicensed spectrum. Unlicensed spectrum is shared by multiple parties and are more prone to interference/contention scenarios. On the other hand, licensed spectrum provides reliable channel allocation for UAV communications and also requires regulatory decisions. The licensed spectrum operations for UAV can be realized via several ways, such as satellite technology, separate licensed spectrum allocated for UAV, or by existing cellular bands. Satellite spectrum is well suited for wide area coverage, but often limited by higher costs, higher latency and lower throughput. Laying out a dedicated spectrum for UAV operations is costly and requires substantial effort to build a system supporting drone operations. To this end, the UAV ecosystem can benefit from existing cellular networks~\cite{fotouhi2019survey} for communication and networking purpose.

\nomenclature{5G/B5G}{Fifth Generation/Beyond Fifth Generation}
\nomenclature{URLLC}{Ultra-Reliable Low Latency Communication}
\nomenclature{mMTC}{Massive Machine Type Communication}
\nomenclature{eMBB}{Extreme Mobile Broadband}
\nomenclature{NR}{New Radio}
\nomenclature{3GPP}{Third Generation Partnership Project}
\nomenclature{RAT}{Radio Access Technology}
\nomenclature{LoS}{Line of Sight}
\nomenclature{NLoS}{Non-Line of Sight}
\nomenclature{BVLoS}{Beyond Visual Line of Sight}
\nomenclature{ATC}{Air Traffic Control}
\nomenclature{HD}{High Definition}
\nomenclature{CNPC}{Control and Non-Payload Communication}
\nomenclature{PER}{Packet Error Rate}
\nomenclature{SOA}{Service Oriented Architecture}

Recently, the ambitious requirements of Fifth Generation and Beyond Fifth Generation (5G/B5G) wireless networks envision to cater to a wider variety of goals in terms of higher coverage and connectivity, ultra-reliable low latency communication (URLLC), support for massive number of devices via machine type communication (mMTC), greater bandwidth and throughput (extreme mobile broadband or eMBB)~\cite{marsch20165g,marsch2016emerging}. The new specifications in Third Generation Partnership Project (3GPP) Rel-15 and improvements for 5G radio interface (termed as 5G New Radio or NR) is designed to offer the above mentioned features~\cite{muruganathan2018overview}. The UAVs are envisioned to be an essential part of 5G/B5G networks with potentials of supporting high data transmission ($\sim$10 Gbits/s), stringent latency (1 ms round trip delay) and enhancements to radio access technologies (RATs). Moreover, the licensed mobile spectrum provides wide accessibility beyond visual line of sight (BVLoS), secure and reliable connectivity enabling cost-effective UAV operation for a multitude of use cases~\cite{li2018uav,bor20195g,naqvi2018drone}.

\subsection{UAV Communication Requirement}
From the communication viewpoint, the requirements of UAV can be classified into two broad categories~\cite{kerczewski2013frequency}:
\begin{itemize}
\item \textbf{Control and Non-Payload Communication (CNPC) -} It refers to the time critical control and safety commands to maintain the flight operations. CNPC includes the navigation, waypoint updates, telemetry report and air traffic control (ATC) updates to ensure secure and reliable UAV operation. CNPC usually demands highly secure and reliable communication with low data rate (few hundred Kb/s) requirements. The reliability requirement for CNPC is less than $10^{-3}$ packet error rate (PER). 
\item \textbf{Payload Communication -} It refers to all the information dissemination activities between UAV and ground station pertaining to a UAV mission. For instance, in a surveillance operation, UAV needs to transmit real time video to the ground station/remote pilot via payload communication. Payload communication demands the underlying transmission medium to be capable of supporting high data rates (often higher in full HD video transmission or wireless backhauling).
\end{itemize}

Table~\ref{commreq} summarizes the rate and latency requirements for UAV cellular communication.

\begin{table}
  \caption{UAV Cellular Communication Requirement~\cite{uav3gpp}}
  \begin{tabular*}{\tblwidth}{@{} CCCC@{} }
   \toprule
    Type & CNPC Uplink & CNPC Downlink & Payload\\
   \midrule
    Rate & $\sim$100 Kbps & $\sim$100 Kbps & $\sim$50 Mbps \\
    Latency & - & $\sim$50 ms & Same as Ground UE \\
   \bottomrule
  \end{tabular*}
  \label{commreq}
\end{table}

%
%

\subsection{Integration Opportunities with Cellular Network}
The integration of UAVs to cellular network falls under three broad paradigms~\cite{zeng2019accessing,vinogradov2019tutorial}, as shown in Fig.~\ref{classification}: 
\begin{itemize}
\item \textbf{UAV-Assisted Cellular Communication -} In this paradigm, UAVs are realized as flying base stations, relays or localization anchors, that can intelligently reposition themselves to assist the existing terrestrial wireless communication system to improve the user perceivable Quality of Experience (QoE), spectral efficiency and coverage gains~\cite{chakraborty2018skyran,sundaresan2018skylite}. This architecture is shown in Fig.~\ref{fig:uav-assisted-cellular}. Due to dynamic mobility and repositioning, the integration of UAV brings many advantages to existing terrestrial communication system~\cite{zeng2016wireless}. The base station mounted on the UAV (flying base station or relays) could be provisioned on demand, which is an absolute appealing solution for disaster management, search and rescue or emergency response. The coverage and data rate of existing cellular networks can be improved by optimal 3D placement and coordination of flying base stations to cater the users need in hotspot areas. These benefits definitely cope well with diverse, dynamic and increasing data demands in 5G/B5G cellular systems. 
\item \textbf{Cellular-Assisted UAV Communication -} This is also known as \textbf{Cellular-connected UAVs}. As shown in Fig.~\ref{fig:cellular-assisted-uav}, flying UAVs are realized as new aerial User Equipments (UEs) coexisting with terrestrial UEs that access the cellular network infrastructure from the sky. This paradigm has gained significant interest in recent times, because of the effective solution for establishing reliable wireless connectivity with ground cellular stations~\cite{zeng2018cellular}.
\item \textbf{UAV-UAV Communication -} In this paradigm, a group of UAVs reliably communicate directly with each other sharing the cellular spectrum with ground users in order to facilitate autonomous flight behaviours, cooperation in a UAV fleets, and collision avoidance. This architecture is shown in Fig.~\ref{fig:uav-uav}. In~\cite{azari2019uav2uav}, the authors investigated reliable and direct UAV-to-UAV communications that leverage same frequency spectrum with uplink of cellular ground users. 
\end{itemize}

In this work, we prioritize the focus on the promising features of cellular-connected UAVs. In the next section, we survey the existing classification works, in order to highlight the key contributions of this work. 

\nomenclature{QoE}{Quality of Experience}
\nomenclature{UE}{User Equipment}
\nomenclature{NFV}{Network Function Virtualization}
\nomenclature{MIMO}{Multiple Input Multiple Output}
\nomenclature{LTE}{Long Term Evolution}
\nomenclature{ANN}{Artificial Neural Network}
\nomenclature{ML}{Machine Learning}
\nomenclature{DPDK}{Data Plane Development Kit}

\renewcommand{\nomname}{\textsc{List of Abbreviations}}
\setlength{\nomitemsep}{-\parsep}
\printnomenclature[2cm]

\section{Related surveys and tutorials}  \label{relatedwork}
There are growing research efforts to investigate the interplay of UAVs with cellular networks. During the last few years, novel solutions have been proposed to solve scientific, technical, socio-economical and security challenges. Several surveys, demonstrations and tutorials are also presented in the literature to provide the unified view of this research domain. These works not only helps the research community to track ongoing research efforts, but also consolidate the necessary knowledge for the interested practitioners and researchers in the community. \\

A majority of the surveys and tutorials mostly focus on the (i) integration opportunity of UAV with 5G/B5G cellular networks from the perspective of UAV-assisted cellular communication or (ii) highlight recent advances, future trends, challenges for UAV cellular communication, or (iii) present detailed analysis and performance study with respect to a specific communication challenge, such as channel modelling, physical layer techniques, security etc. However, our work aims at focussing on the paradigm of cellular-assisted UAV communications. In order to emphasize the relevance and uniqueness of our current survey work compared to existing surveys, first, we plan to summarize the existing surveys and tutorials along with works pertaining to cellular-connected UAVs in Table~\ref{relworks}. \\



As summarized in Table~\ref{relworks}, majority of existing surveys are based on UAV-assisted cellular communication and discusses them in-depth. There are few surveys that focuses on cellular-connected UAV paradigm, but these existing works are largely fragmented and do not provide a holistic view of this paradigm. In other words, only few selected aspects of cellular-connected UAV like UAV-ground channel modelling or trajectory optimization or MIMO are studied in depth so far. These works do not present an extensive study including all kinds of research highlights dedicated to cellular-connected UAVs, rather present a singular topic in depth. Thus, a unified work providing the broad picture of all kinds of research developments is still missing. \\

With this survey, we aim at addressing this gap and focus solely on cellular-connected UAVs. The research highlights pertaining to state-of-the-art advancements, synergistic integration challenges of UAVs as aerial users in 5G/B5G cellular networks, underlying network architectures, physical layer enhancements of 5G, field trials, simulations and testbed developments are some of unique contributions made in this survey. The cloudification and softwarization of network resources portrayed as the interplay of Network Functions Virtualization (NFV) and cloud computing technologies for the cellular-connected UAVs is also presented from the architectural context of enabling 5G/B5G innovations supporting them.

\subsection{Key Contributions}
The key contributions of this work are the following. The final column of Table~\ref{relworks}, bearing the heading ``This Work", also summarizes the contributions made in this work:
\begin{itemize}
\item To present an overview of emerging applications and taxonomy of use cases for cellular-connected UAVs;
\item To highlight the state-of-the-art trends of communication requirements of UAVs and detailed discussion of design challenges, which must be accounted for successful integration of this technology within 5G/B5G cellular systems;
\item To showcase the emerging 5G technology innovations in network architectures such as virtualization \& softwarization of network resources, slicing \& physical layer improvements in the interest of cellular-connected UAVs;
\item To present the detailed efforts for design and development of experimental testbeds, trials and prototyping carried out by academia, industries and standardization bodies to understand the gap of theoretical analysis and realistic deployments;
\item To identify a fairly exhaustive outline of features for realization of an ideal experimental prototype for cellular-connected UAV \& existing works to achieve them;
\item To discuss about the ongoing standardization activities, regulatory frameworks, market and socio-economic issues that must be thoroughly investigated before successful and widespread adoption of cellular-connected UAVs;
\item To present insights to future research opportunities.
\end{itemize}

\begin{figure}[h]
\centering
\includegraphics[width=1\linewidth]{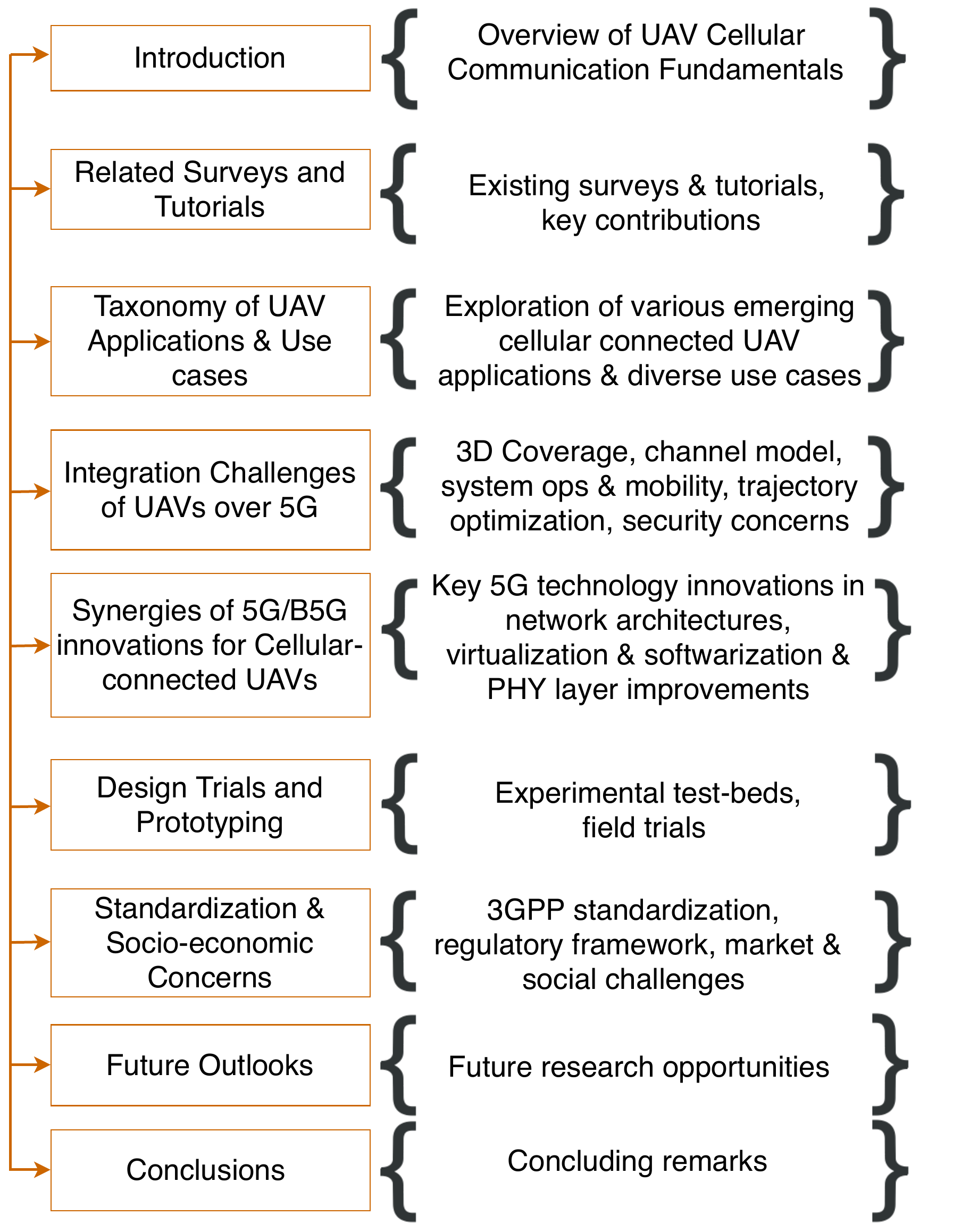}
\caption{High level organization of this work}
\label{org}
\end{figure}

The high level organization of this work is summarized in Fig.~\ref{org}.

\begin{landscape}
\begin{table}[t]
\small
\centering
\caption{Existing surveys and tutorials for UAV cellular communication}
\setlength{\tabcolsep}{2.5pt} 
\renewcommand*{\arraystretch}{1.8} 
\begin{tabular*}{21.58cm}{|c|c|c|c|c|c|c|c|c|c|c|c|c|c|c|c|c|c|c|c|c|c|}
\hline 
\textbf{\parbox{2cm}{Broad\\Category}} & \textbf{\parbox{2.5cm}{References \contour{black}{$\rightarrow$}\\Contributions \contour{black}{$\downarrow$}}} & \textbf{~\cite{mozaffari2019tutorial}} & \textbf{~\cite{fotouhi2019survey}} & \textbf{~\cite{zeng2019accessing}} & \textbf{~\cite{shakhatreh2019unmanned}} & \textbf{~\cite{khuwaja2018survey}} & \textbf{~\cite{Khawaja2019}} & \textbf{~\cite{gupta2015survey}} & \textbf{~\cite{zeng2016wireless}} & \textbf{~\cite{mkiramweni2019survey}} & \textbf{~\cite{yan2019comprehensive}} & \textbf{~\cite{vinogradov2019tutorial}} & \textbf{~\cite{li2018uav}} & \textbf{~\cite{motlagh2016low}} & \textbf{~\cite{hayat2016survey}} & \textbf{~\cite{azari2018reshaping}} & \textbf{~\cite{geraci2018uav}} & \textbf{~\cite{azari2019cellular}} & \textbf{~\cite{wang2018network}} & \textbf{~\cite{zeng2018cellular}} & \textbf{\textcolor{blue}{\parbox{1cm}{\vspace{.20\baselineskip}This\\Work\vspace{.20\baselineskip}}}} \\ 
\hline
\multirow{2}{*}{\parbox{2cm}{Nature of\\Integration}} & 
\parbox{2.5cm}{\vspace{.25\baselineskip}UAV-Assisted \\Cellular Comm\vspace{.25\baselineskip}} &  \checkmark &  \checkmark & \checkmark &  \checkmark  &  \checkmark & \checkmark   &   &  \checkmark  & \checkmark  &   & \checkmark  &  \checkmark  & \checkmark  &  \checkmark  &  \checkmark  &   &   &    &   & \textcolor{blue}{\checkmark} \\ \cline{2-22}  &
\parbox{2.5cm}{\vspace{.25\baselineskip}Cellular-Assisted \\UAV Comm\vspace{.25\baselineskip}}   &  \checkmark &   &  \checkmark &  \checkmark   &   \checkmark &   \checkmark  &   &    &   &    &  \checkmark  &   \checkmark  &  \checkmark  &   \checkmark  &   &  \checkmark  &  \checkmark  &   \checkmark  &  \checkmark  & \textcolor{blue}{\checkmark} \\
\hline
\multirow{1}{*}{\parbox{2cm}{Applications\\\& Use cases}} & \parbox{2.5cm}{\vspace{.25\baselineskip}Applications \\\& Use cases\vspace{.25\baselineskip}}   &  \checkmark &   \checkmark &   &  \checkmark  &   &    &   &    &   &    &   &    &   &   \checkmark  &   &   &   &   &   & \textcolor{blue}{\checkmark} \\
\hline
\multirow{4}{*}{\parbox{2cm}{Design \&\\Challenges}} & \parbox{2.5cm}{\vspace{.25\baselineskip}Technical \\Challenges\vspace{.25\baselineskip}}   & \checkmark  &   & \checkmark  &  \checkmark  &  \checkmark  &  \checkmark  &   \checkmark &  \checkmark  &  \checkmark  & \checkmark & \checkmark  &    \checkmark &  \checkmark &  \checkmark  &  \checkmark &  \checkmark & \checkmark  &  \checkmark  & \checkmark  & \textcolor{blue}{\checkmark} \\  \cline{2-22}  &
\parbox{2.5cm}{\vspace{.25\baselineskip}Propagation \\Channel Models\vspace{.25\baselineskip}}   & \checkmark  &   & \checkmark  & \checkmark   & \checkmark  & \checkmark   &   & \checkmark   & \checkmark  & \checkmark   &  \checkmark &  \checkmark  &   &  \checkmark  & \checkmark  &   & \checkmark  & \checkmark   & \checkmark  & \textcolor{blue}{\checkmark}  \\   \cline{2-22}  &
\parbox{2.5cm}{\vspace{.25\baselineskip}Mobility \& \\Handovers\vspace{.25\baselineskip}}   &  &   &   &    &   &    &  \checkmark &    &   &    &   &    &   &    &   &   &   &    &   & \textcolor{blue}{\checkmark} \\  \cline{2-22}  &
\parbox{2.5cm}{\vspace{.25\baselineskip}Trajectory \\Optimization\vspace{.25\baselineskip}}  &  \checkmark &   & \checkmark  &    &   &    &   &    &   &    &   &    &   &    &   &   &   &    &   & \textcolor{blue}{\checkmark} \\ 
\hline
\multirow{4}{*}{\parbox{2cm}{Technology \&\\ Experiment}} & \parbox{2.5cm}{\vspace{.25\baselineskip}Network \\Architecture\vspace{.25\baselineskip}}   &  &  &   &    &    &   & \checkmark  &   &    &   &    &  \checkmark  & \checkmark  &   &    & \checkmark  &   &  \checkmark  &   & \textcolor{blue}{\checkmark} \\  \cline{2-22}  &
\parbox{2.5cm}{\vspace{.25\baselineskip}5G/B5G \\Innovations\vspace{.25\baselineskip}}   &  &   & \checkmark & \checkmark   & \checkmark & \checkmark   & \checkmark  &    &   &  \checkmark  &   &  \checkmark  &   &    &   &   & \checkmark  & \checkmark   &  \checkmark  & \textcolor{blue}{\checkmark}  \\  \cline{2-22}  &
\parbox{2.5cm}{\vspace{.25\baselineskip}Experimental\\Prototyping\vspace{.25\baselineskip}}   &  &  \checkmark  &  &    &   &    &   &    &   &    &   &    &   &    &   &   &   &    &   & \textcolor{blue}{\checkmark}  \\  \cline{2-22}  &
\parbox{2.5cm}{\vspace{.25\baselineskip}Ideal Features\\of Prototype\vspace{.25\baselineskip}}   &  &   &  &    &   &    &   &    &   &    &   &    &   &    &   &   &   &    &   & \textcolor{blue}{\checkmark}  \\ 
\hline
\multirow{3}{*}{\parbox{2cm}{Harmonization\\\& Compliance}} &
\parbox{2.5cm}{\vspace{.25\baselineskip}Standardization\vspace{.25\baselineskip}}   &  & \checkmark  &  &    &   &    &   &    &   &    &   &    &   &    &   & \checkmark  &   &    &   & \textcolor{blue}{\checkmark} \\  \cline{2-22}  &
\parbox{2.5cm}{\vspace{.25\baselineskip}Regulations\vspace{.25\baselineskip}}   &  \checkmark &  \checkmark  &  &    &   &    &   &    &   &    &   &    &   &    &   &   &   &    &   & \textcolor{blue}{\checkmark} \\  \cline{2-22}  &
\parbox{2.5cm}{\vspace{.25\baselineskip}Communication\\Requirement \vspace{.25\baselineskip}}   &   &   & \checkmark  &    &   &    &   &  \checkmark  &   &    &   &    &  \checkmark & \checkmark  &   &   &   &    & \checkmark  & \textcolor{blue}{\checkmark} \\ 
\hline
\multirow{3}{*}{\parbox{2cm}{Socio-economic\\Concerns}} &
\parbox{2.5cm}{\vspace{.25\baselineskip}Security Aspects\vspace{.25\baselineskip}}   &  &  \checkmark  & \checkmark  &    &   &    &   &    &   &    &   &    &   &    &   &   &   &    &   & \textcolor{blue}{\checkmark} \\  \cline{2-22}  &
\parbox{2.5cm}{\vspace{.25\baselineskip}Social Concerns\vspace{.25\baselineskip}}   &  & \checkmark  &  &    &   &    &   &    &   &    &   &    &   &    &   &   &   &    &   & \textcolor{blue}{\checkmark} \\  \cline{2-22}  &
\parbox{2.5cm}{\vspace{.25\baselineskip}Market Concerns\vspace{.25\baselineskip}}   &  &   &  &    &   &    &   &    &   &    &   &    &   &    &   &   &   &    &   & \textcolor{blue}{\checkmark} \\
\hline
\end{tabular*}
\label{relworks}
\end{table} 
\end{landscape}

First, we begin with the detailed taxonomy of application domains and corresponding use cases for cellular-connected UAVs, and then highlight the key integration challenges of UAVs being supported from cellular 5G/B5G systems. For seamless integration, the recent technical innovations of 5G/B5G mobile network architectures and physical layer improvements are also presented. Then, we highlight the testbeds, field trials and measurement campaigns that showcase some early efforts to develop working prototypes of cellular-connected UAV. Furthermore, the ongoing standardization works, regulatory and socio-economic concerns are also discussed that must be accounted before successful adoption of this new technology.

\section{Taxonomy of UAV Applications and Use cases} \label{background}
\begin{figure*}[t]
\centering
\includegraphics[width=0.94\linewidth]{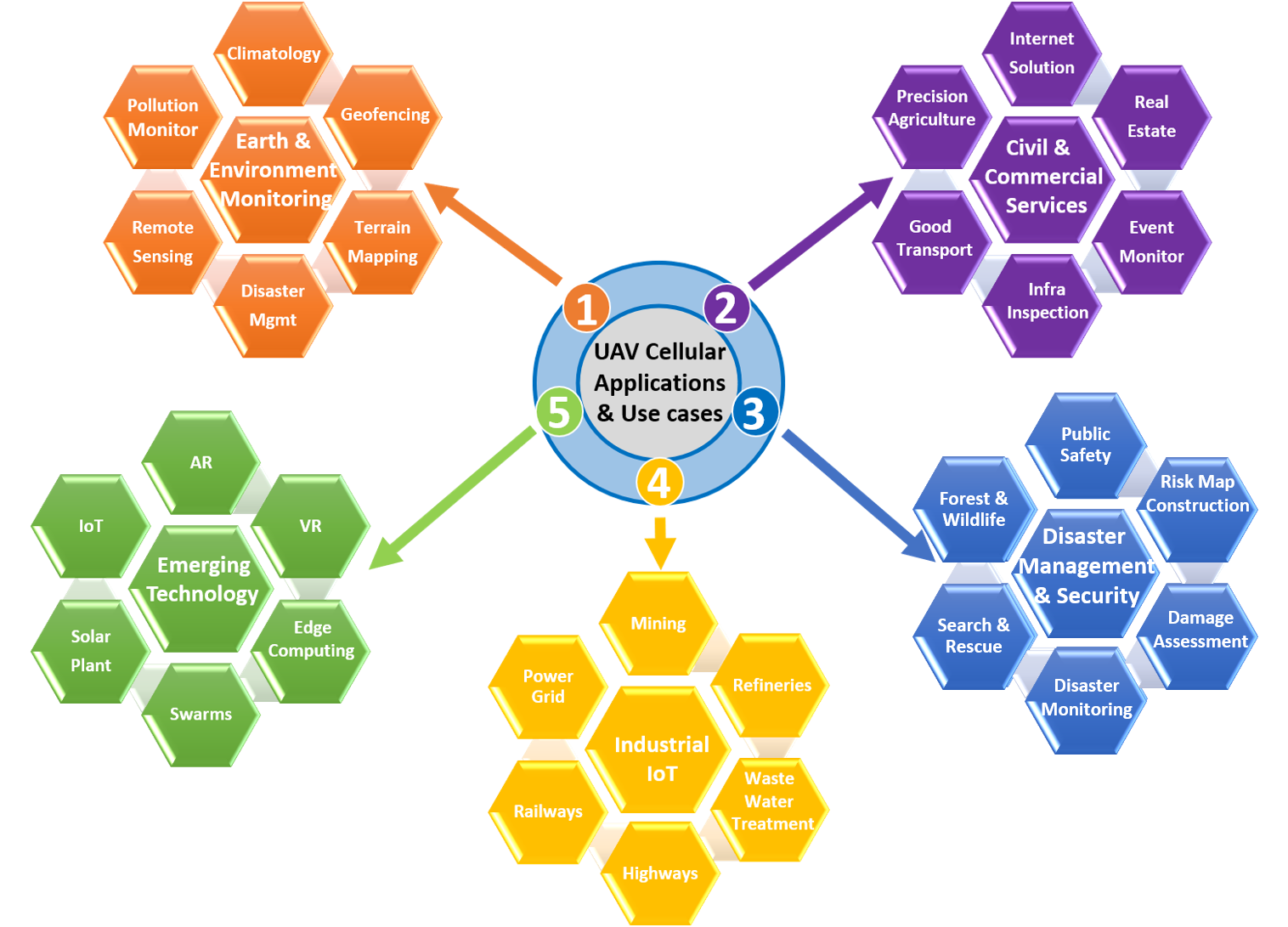}
\caption{Taxonomy of UAV Cellular applications \& use cases}
\label{taxonomy}
\end{figure*}

Cellular-connected UAVs find their applicability in a wide range of emerging applications with varying demands and goals. In this work, we showcase some of the attractive researched domains as a starting basis for the following discussion. A bird's-eye view of this section is presented in Fig~\ref{taxonomy}.


\subsection{Earth and Atmospheric Observations}
As an innovative and efficient platform for gathering data, UAVs have become a preferred choice over traditional geomatics mechanisms of data acquisition. UAVs could autonomously fly in a defined trajectory and could precisely capture real-time measurements of the ongoing geophysical processes for abnormal hazards, such as volcanoes, landslides, sea dynamics, earthquakes, etc. Furthermore, the UAVs are equipped with various sensors to capture atmospheric temperature, pollutant levels in the air, carbon emissions, terrestrial biomass characterization, precipitation distribution in industrial zones, etc. As an efficient mechanism, the deployment of a fleet of UAVs, equipped with onboard sensors can perform the sensing for the presence of pollutant levels or any hazardous chemicals in the target areas~\cite{alvear2017using,alvear2018}. 

In a disaster situation, first 48 to 72 hours are very crucial to perform any kind of mitigation to the damage or outage and to restore the normal state of the environment. The response time is the key in saving lives in the affected regions. The major problems in these initial hours are: lack of proper communication infrastructure, massive or often unpredictable losses of lives and property. Thus, the situation forces the first responder teams to implement and improvise the search and rescue (SAR) mission to be conducted quickly and efficiently. Latest advancements of UAVs and sensor networks are capable to meet this need in terms of disaster prediction, assessment and fast recovery. UAVs can gather the information (e.g., situational awareness, early warnings, persons movement) during disaster phase and these information are helpful for first responder teams to react efficiently. UAVs can re-establish the communication infrastructure (\emph{i.e.,} UAV-assisted paradigm) destroyed at the time of disaster.


\subsection{Civil and Commercial Services}
Government constructions and public infrastructures such as highways and railways are greatly benefited by these flying platforms for efficient surveillance, land surveying, tracking workers and employees, on-site construction and demolition~\cite{shakhatreh2019unmanned,erdelj2019uavs,trotta2018uavs}. Furthermore, UAV-based delivery systems are gaining wide popularity in logistics domain to achieve faster and cost effective good delivery service~\cite{grippa2019drone}. Such a system handles consumer orders, manages autonomous flight and status tracking using real time control. Google’s Wing project and Amazon Prime Air are the some of the efforts to realize such a use case of UAVs. In precision agriculture, UAVs are capable of observing the agricultural fields for health monitoring, spraying pesticides and perform hyper spectral imagery. Such activities by humans are time consuming and prone to risks. Unmanned aircrafts are well suited for such use cases enhancing productivity and cost efficacy. The cellular operators have started envisioning UAVs as backup wireless infrastructure (flying base stations or relays) in the absence of terrestrial communication infrastructure to boost network capacity~\cite{lin2018sky}. Google’s Loon project aims to provide ubiquitous Internet services \& wireless connectivity to both remote and rural areas by employing high altitude platform (HAP) UAVs as balloons.

\subsection{Disaster management \& Security}
UAVs are an effective means of surveillance and monitoring of areas stricken by a natural disaster. For instance, autonomous UAVs are sent to landslide, fire, earthquake and flooding areas to help with assessing the risks, the damages and support first responders teams as well as providing connectivity to isolated people \cite{erdelj2017wireless,erdelj2017help,erdelj2018}. Similarly, low cost UAVs revolutionize the conservation and management of forest and wildlife ecosystem by assisting in counting animal populations, tracking illegal activities, etc.
UAVs are also an effective means of surveillance and control for the homeland security and public safety~\cite{motlagh2017uav,boccardo2015uav,he2017drone}. In case of anti-terroristic operations, UAVs are used to develop and prepare for situational awareness of threat, carrying out pre-emptive strikes or reconnaissance mission. UAVs assist in speeding up the rescue and recovery (search and rescue) missions in certain disastrous and crime control situations in a target area. 


	
\subsection{Industrial IoT platforms (IIoT)}
Industry 4.0 is an emerging paradigm embracing next generation industrial developments with the ideas of using Internet of Things (IoT) to industrial automation, cyber physical systems, smart production and service systems. This industrial revolution is a gateway to boost economy and operational excellence under the umbrella term of ``Smart Factory”. 

UAVs have already begun to become a vital component of Industrial IoT platforms~\cite{salhaoui2019smart,lagkas2018uav}. Practical usage of UAVs in industrial settings include monitoring terrains of manufacturing sites or regions that are impenetrable for humans due to hazardous exposures. The manual on-site inspection carried out by humans are time-consuming and often include very challenging terrains with inaccessible/unsafe zones. Such human-driven inspections pose threats to human lives. On the bright side, not only industrial UAVs can penetrate complex and inaccessible areas, but also are equipped with a multitude of sensors with cognitive computing to facilitate on-demand real-time bidirectional communication with industrial control stations. UAVs used in industrial settings can measure many parameters for the region under study via onboard sensors, such as electric and magnetic field strength, humidity, temperature, pressure in the atmosphere, methane or toxic pollutants. The communication could occur the same way as an IoT sensor sending signals to the Supervisory Control and Data Acquisition (SCADA) system.
	  
	  
\subsection{Emerging Technologies}
Some emerging technologies such as augmented reality (AR) and virtual reality (VR) combined with capabilities of UAV open up novel possibilities~\cite{chakareski2017aerial}. Real life videos from high altitude or high quality aerial photographs bring a great look and feel experience for users. Also, in the enterprise markets, the VR technology clients can accelerate buyer's decisions by presenting them best scenery and viewing of the real estate. AR- and VR-enabled UAVs are also used for virtual tour of the real environments, 3D models of buildings, graphical overlays of maps, streets,  gaming, etc.


\nomenclature{LAP}{Low Altitude Platform}
\nomenclature{HAP}{High Altitude Platform}
\nomenclature{SAR}{Search and Rescue}

\subsection{Consolidated summary of lesson learnt and rationale of this work}
The important lessons learnt in the previous Sections can be summarized in the following two main items:
\begin{itemize}
\item The popularity of UAV is growing day-by-day and it is considered as a preferred technology to cater to a wide variety of emerging real-world use cases. UAVs can be autonomous, intelligent, adaptive and highly mobile. From communication and networking perspective, UAVs play an important role in cellular domain. The cellular ecosystem can benefit from UAV technology. UAVs can be efficiently integrated to existing cellular networks as a flying base station or a relay or an aerial UE. These different types of integration showcase several promising applications and use cases.
\item Owing to the implicit benefits of cellular networks in terms of ubiquitous accessibility, large coverage, scheduled and safe information exchange protocols, cellular-connected UAVs are well suited and find their applicability in many real-world applications such as earth and environmental observation, civil infrastructure and surveillance, defence and security, industrial IoT platforms, etc. Integrating UAVs to 5G/B5G cellular systems proves to be a win-win situation for both the parties.
\end{itemize}

This work aims at presenting an extensive study of cellular-connected UAVs, where the UAVs are integrated into the existing cellular networks as new aerial UEs. In order to carry out a mission specific task, UAVs require support from ground infrastructure (base station/control station), with which they exchange CNPC commands in downlink direction, and both CNPC \& payload data in uplink. 
Cellular-connected UAVs bring several open challenges and operational complications that need to be thoroughly investigated and motivate our work to offer researchers and practitioners a handful guide to approach this field.


\section{Integration Challenges of UAVs over 5G} \label{uav5g}
The aerial communications and networking of cellular-connected UAVs pose several challenges to thoroughly investigate. For instance, a reliable and low latency communication for efficient control of the UAV is of utmost importance. Existing cellular infrastructures are primarily designed and developed to offer enhanced communication services for the terrestrial users. Also, the geographical terrains with limited coverage from terrestrial Base Station (BS) may not provide the required connectivity services to the cellular-connected UAVs, thereby demand promising solutions for successful adoption of this technology.

Various studies and research efforts have shown tremendous potential for the support and operation of low altitude UAVs using cellular networks~\cite{ullah20195g}. The benefits of cost-effective cellular spectrum in terms of low latency and high throughput connectivity services, make it a suitable candidate for integration of UAVs. Moreover, this technology is scheduled, robust, secure and offers reliable services. In terms of the security aspects of data communication, existing mobile networks already encompass the needful security and authentication features in their protocol layers. A work item to study and evaluate LTE as a potential candidate for UAV operation is carried out in 3GPP Rel-15, and the results are summarized in TR 36.777~\cite{muruganathan2018overview}. In addition to existing cellular spectrum bands (600 MHz - 6 GHz), 5G ecosystem is also considering the use of spectrum in millimeter wave (mmWave) bands (24-86 GHz). As a foundation of cellular operations, the licensed spectrum provide scheduled, reliable and wide area connectivity that can potentially be leveraged for UAV operations in BVLoS range.

There are a lot of challenges to be tackled in order to make the cellular-connected UAVs as an attractive solution for a plethora of emerging use cases. In the following subsections, we highlight the primary design challenges and perspectives to be considered for cellular-connected UAVs, as well as the studies and solutions already available. 

\begin{figure}[b]
\centering
\includegraphics[width=1\linewidth]{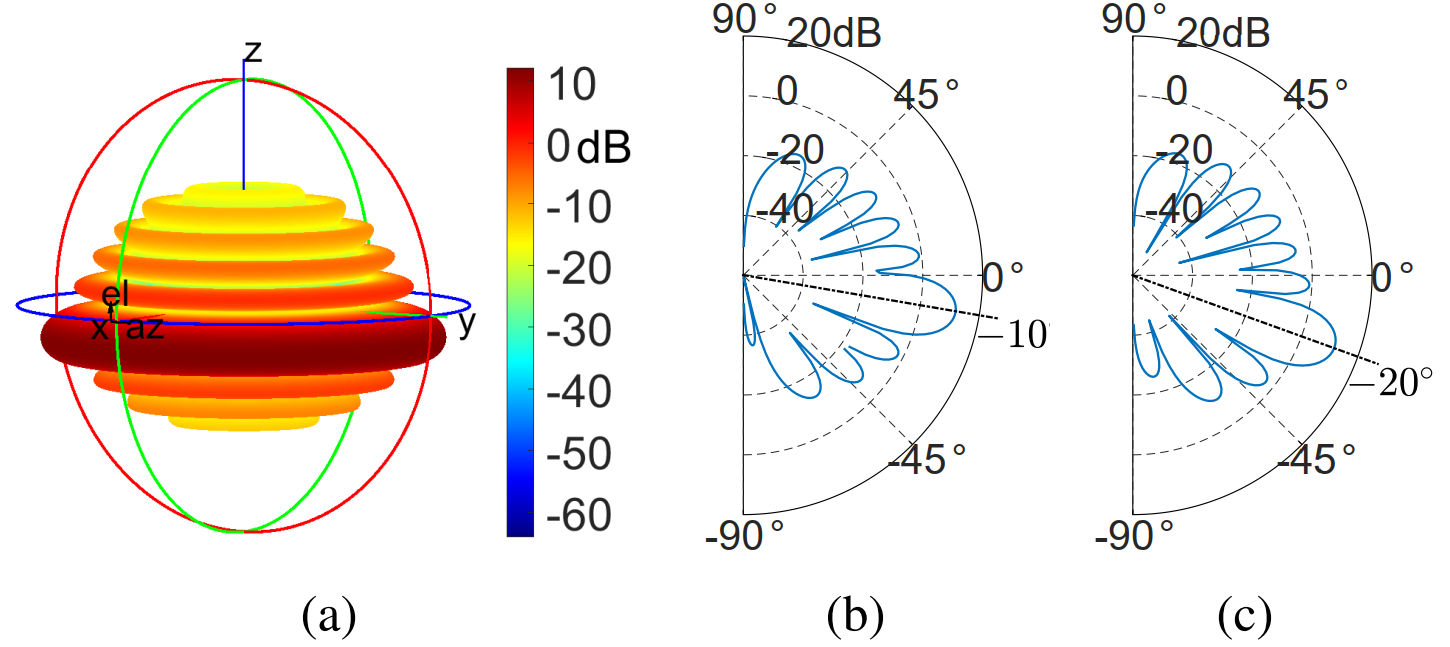}
\caption{Power gain and elevation pattern of ground BS~\cite{Lyu2019}}
\label{pwgain}
\end{figure}

\subsection{Three Dimensional (3D) Coverage Model}
\subsubsection{Preliminary}
The existing radio access technologies are not primarily suited for supporting flying radio devices as their deployments are mainly focused to optimally serve the ground UEs (or terrestrial UEs). The base stations (eNodeBs) are typically designed and developed to provide optimal performance to the ground users. The current eNodeBs are downtilted to serve above purpose. Down-tilting the antennas produces radiation patterns that are not useful to serve aerial UEs, which are expected to be positioned at different altitudes with respect to the ground surface. The inherent assumptions made for the ground UEs are quite different from aerial UEs. The aerial users typically fly higher than the BS antenna height and therefore, need 3D coverage suitable for varying UAV altitude~\cite{mozaffari20183d}. The BS antennas of LTE networks may provide weak channel gain by using their antenna side lobes. In the 3D space, the coverage criterion of UAVs are functions of BS antenna height, UAV altitude, antenna pattern, and association rules. Hence, the network model for aerial users coexisting with ground users necessitates a 3D coverage model~\cite{xu2019cellular}.

\subsubsection{Associated Works and Illustrative Results}
In~\cite{Lyu2019}, the authors present 3D coverage and channel modelling of cellular-connected UAVs in the downlink and uplink directions. The BS antenna pattern tremendously impacts the coverage distribution that affects the UAV operation and mobility. As the down-tilt angle increases, the ground BS offers smaller gains to UAV above the BS height, thereby impacting the uplink and downlink coverage probabilities.  In Fig.~\ref{pwgain} (a), the power gain pattern is illustrated for a synchronized uniform linear antenna array (ULA) with 10 co-polarized dipole antenna elements and BS down-tilt angle $ \theta_{tilt} = -10\si{\degree}$. (b) and (c) are the 2D elevation pattern for BS with down-tilt angle for $ \theta_{tilt} = -10\si{\degree}$ and $ \theta_{tilt} = -20\si{\degree}$, respectively. 

\subsection{UAV-Ground Channel}
\subsubsection{Preliminary}
One of the primary design challenges in realizing the cellular-connected UAVs is to ensure harmonious coexistence mechanisms between ground users and aerial users~\cite{azari2017coexistence}. Proper UAV-ground interference management is central to realize this coexistence. Also, the interference patterns in ground BS to UAVs communication link experiences remarkable difference than that of link between ground BS to ground UE~\cite{khuwaja2018survey}. The higher altitude of UAVs than base stations results in LoS links, which are more reliable than the link with ground users. Additionally, they exploit large macro diversity gains being served from several BSs. On the other hand, the dominant LoS links create more uplink/downlink interference as compared to ground users, thereby making the interference management (ICIC) highly difficult. Other relevant effects to take into account are fading, shadowing and path-loss. Existing ICIC mechanisms may be well suited for current cellular designs, but fail to handle UAV interference management, which involves many BSs and impose limitations due to high complexity. 

Therefore, there is a need for efficient interference management techniques for harmonious coexistence of ground users and UAVs. There are several works in literature~\cite{mei2019cellular,zhou2018coverage,xu2019cellular} that investigate this problem considering downlink and uplink interference. 

The communication channel mainly involves two types of links, namely Ground-to-UAV (G2U) link and UAV-to-Ground (U2G) link. In cellular-connected UAV, the G2U link serves the downlink purpose of control and command for proper UAV operations, whereas U2G link serves the uplink purpose of payload communication. Rayleigh fading is the commonly used small-scale fading model for terrestrial channel model, whereas due to the presence of LoS propagation characteristics, Nakagami-m and Rician small-scale fading are usually preferred for U2G channels. The large-scale fading is affected because of the 3D coverage region and varying altitude of UAV. The large-scale fading models used can be based on a free-space channel model or altitude/angle dependent channel model or probabilistic LoS models:

\begin{itemize}
\item Free-space model - In free-space channel model, there is no effect of fading and shadowing with very limited obstruction. This model is typically suited for rural regions where the LoS assumption holds valid between high altitude UAVs and ground station. However, in urban environment, the low altitude UAVs may encounter non-LoS links, therefore need other approaches to properly map with the propagation environment. 
\item Altitude/Angle dependent model - In this case, the channel parameters such as shadowing and path loss exponents are functions of UAV altitude or elevation angle. These models find their applicability in urban or sub-urban regions depending upon the deployment. However, if the altitude does not change or UAVs fly horizontally, altitude dependent models may not be found suitable. The elevation angle based models are mostly used for theoretical study purpose and existing literatures are also limited in this regard.
\item Probabilistic LoS model - The models based on this approach are typically suited for urban environment where the LoS and NLoS link between UAV and ground are considered, due to buildings, obstacles or blockages. Moreover, the LoS and NLoS components are separately modelled based on their occurrence probability in urban environment. The nature of urban environment with respect to building heights and density are key factors that statistically determine the LoS and NLoS propagation characteristics.
\end{itemize}

\subsubsection{Associated Works and Illustrative Results}
The study item of 3GPP TSG on the enhanced LTE support for aerial vehicles~\cite{uav3gpp} highlights the channel modelling between ground base station and UAV flying at different altitudes. The study includes the modelling of small scale fading, path loss, shadowing and LOS probability ($P_{los}$) for three 3GPP deployment scenarios, namely Urban-Micro (UMi), Urban-Macro (UMa) and Rural-Macro (RMa). 
The LoS probability is specified by: 
\begin{itemize}
\item 2D distance between UAV and ground station ($d$)
\item Altitude of UAV ($h_{u}$)
\end{itemize}

The existing terrestrial communication channel model can be directly used for low UAV altitude (height below certain threshold $H_{low}$) to model the LoS probability. For altitude greater than a certain threshold $H_{high}$, 3GPP suggests to use 100\% LoS probability. For height in between $H_{low}$ and $H_{high}$, the LoS probability is a function of $d$ and $h_{u}$. Hence, for the three deployment scenarios, $P_{los}$ is given by,

\[
    P_{los}= 
\begin{cases}
    UE\_P_{los}, & \text{if } 1.5  \text{ meter } \leq h_{u} \leq H_{low}  \\ 
    f(h_{u}, d), & \text{if } H_{low} \leq h_{u} \leq H_{high}  \\ 
    1, & \text{if } h_{u} \geq H_{high} \text{ and } h_{u} \leq 300 \text{ meter }
\end{cases}
\]
$UE\_P_{los}$ is the LoS probability for ground mobile terminal in conventional terrestrial communication in Table 7.4.2 of~\cite{channelfreq}. $f(h_{u}, d)$ is given by,
\[
    f(h_{u}, d) =
\begin{cases}
    1, & \text{if } h_{u} \leq l1 \\
    \frac{l1}{h_{u}} + exp(\frac{-h_{u}}{p1})(1 - \frac{l1}{h_{u}}) , & \text{if } h_{u} > l1 \\
\end{cases}
\]

The variables $l1$ and $p1$ are given as the logarithmic increasing function of UAV height $h_{u}$ as specified in~\cite{uav3gpp}. The values of $H_{low}$, $H_{high}$, $p1$ and $l1$ are also defined with respect to different 3GPP deployment scenarios. Table B-2 and B-3 in~\cite{uav3gpp} provides detailed path-loss and shadowing standard deviation, respectively.

\subsection{System Operations \& Mobility}
\subsubsection{Preliminary}
UAVs are inherently mobile in nature and section~\ref{background} highlights many use cases of cellular-connected UAVs that implicitly demand BVLoS~\cite{stanczak2018mobility}. The mobility and handover characteristics of terrestrial cellular users are quite different from the 3D aerial mobility of cellular-connected UAVs. With increase in height, the radio environment changes and mobile UAVs face connectivity challenges. In this case, the performance of the system depends on the handover rates, including failed and successful handovers and radio link failures. Radio link failures occurs when the UAV is unable to maintain a successful connection with the serving cell. This could be because of the problematic RACH or expiry of timers or after a certain maximum number of retransmissions is reached~\cite{arshad2016handover}.

In cellular-connected UAV, the protocol operations and regulatory needs of UAVs as aerial users are quite different from the ground user. Hence, the network must first detect if the user device is aerial or not~\cite{zhang2019research}. This detection can be driven by the ground BS by estimating:
\begin{itemize}
\item the elevation angle of the reference signal;
\item vertical location (altitude) or velocity of user device;
\item path loss/delay spread measurement of user devices.
\end{itemize}

\subsubsection{Associated Works and Illustrative Results}
\begin{figure}[t]
\centering
\includegraphics[width=1\linewidth]{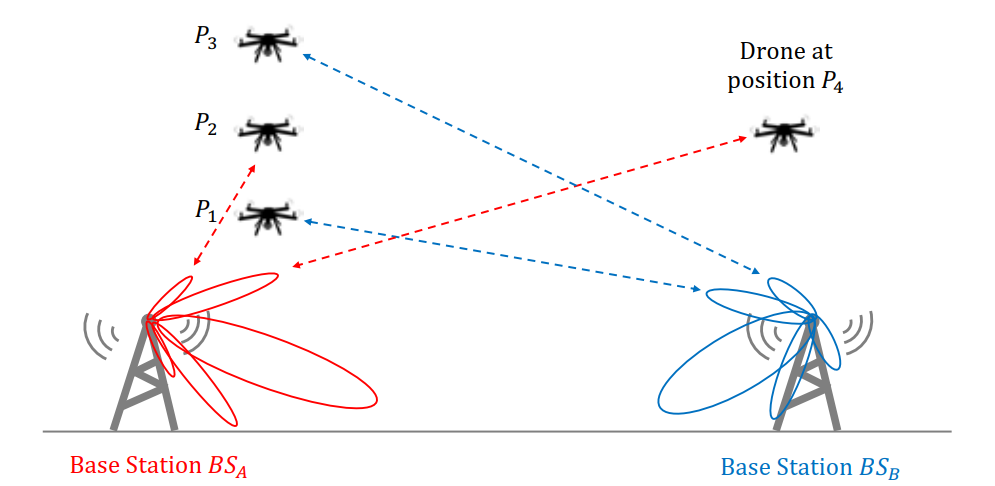}
\caption{UAVs being served from side lobes~\cite{fakhreddine2019handover}}
\label{handover1}
\end{figure}

The handover characteristics vary significantly between ground UEs and aerial UE due to the nature of cell selection, as shown in Fig.~\ref{handover1}. In~\cite{fakhreddine2019handover}, the authors demonstrated the impact of UAV flight path on handovers. The results show that UAVs are prone to frequent handovers, and ping-pong handovers, due to varying altitude and speed. Even smaller flight distances can have a large impact on handover rate. Also, the handover frequency increases when flight altitude increases. Table~\ref{ho} summarizes the number of handovers occurring per minute for UAV, as compared to terrestrial users. Scenario1 is equivalent to a ground user having one handover per minute. However, in scenario4, UAVs, at an altitude of 150m, experiences 5 handovers per minute. Many of the handovers are unnecessary and generate high signalling overhead. Handover decisions are mainly made depending upon received RSRP (Referenced Signal Referenced Power) values from different BS antennas. Ground users are benefited by this approach, because the radio transmission power are directed to ground from the main lobes of the antenna, thereby improved radio power and every received RSRP is well separated from others. However, the aerial users are served primarily by the antenna side lobes, whose RSRP tends to be very similar to the radio power from other surrounding BS. Hence, the UAV connects with more cells (distant cells), as there is a small difference in the RSRP values resulted from BS antenna side lobes. 

Hence, integration of cellular-connected UAVs with future 5G/B5G networks necessitates enhanced solutions for cell selection and handovers that seamlessly cover changing altitudes of UAVs and support their 3D mobility patterns.


\begin{table}[pos=t]
  \caption{Rate of handovers with varying UAV altitude~\cite{fakhreddine2019handover}}
  \begin{tabular*}{\tblwidth}{@{} CCC@{} }
   \toprule
    Scenario & Height (Meters) & \#Handovers/Minute \\
   \midrule
    1 & 10 & 1.0 \\  
    2 & 50 & 1.9    \\
    3 & 100 & 4 \\  
    4 & 150 & 5    \\
   \bottomrule
  \end{tabular*}
  \label{ho}
\end{table}


\begin{figure}[b]
\centering
\includegraphics[width=0.7\linewidth]{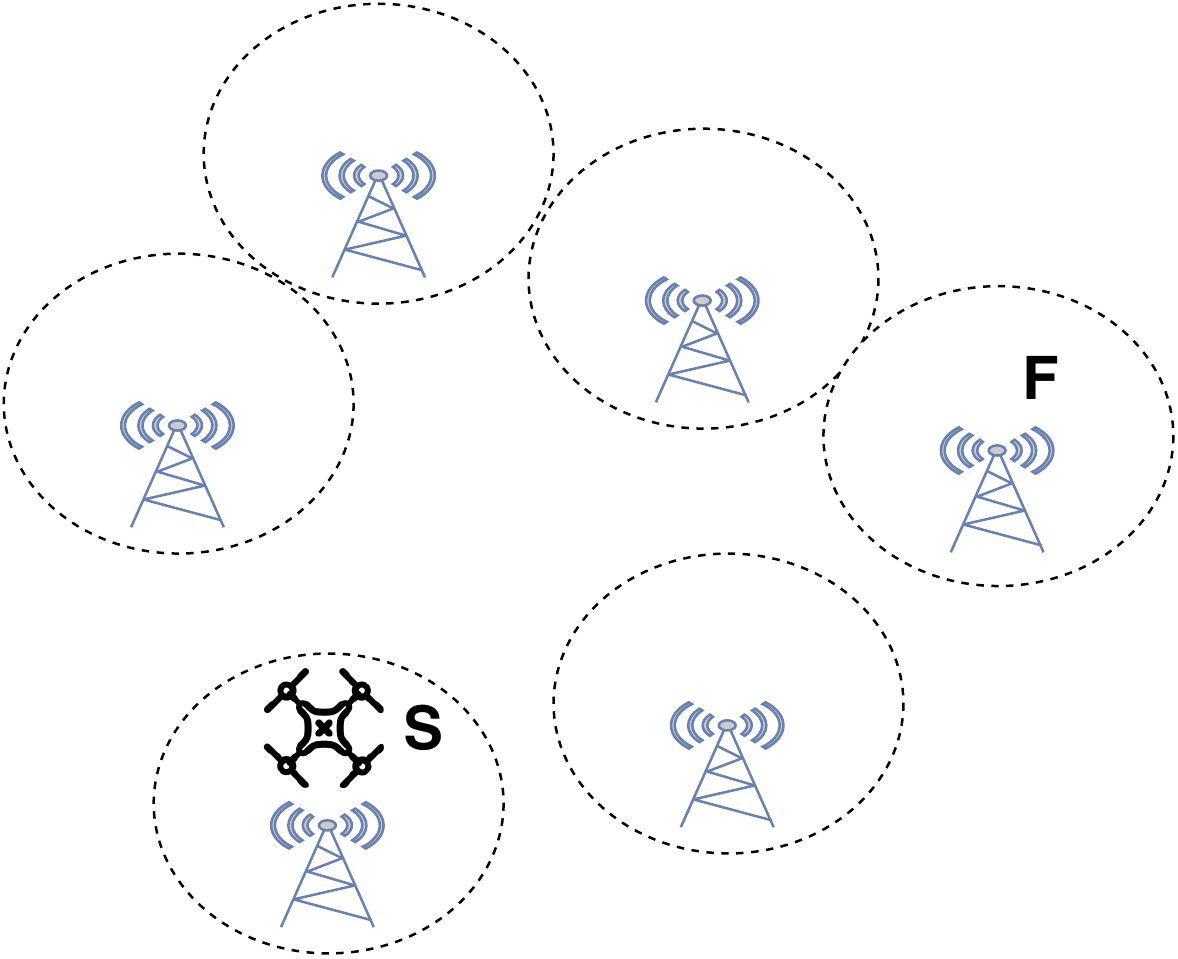} 
\caption{UAV trajectory with cellular discontinuity}
\label{trajectory}
\end{figure}

\subsection{Trajectory Optimization}
\subsubsection{Preliminary}
UAV trajectory or flight path refers to the path through which UAV completes its mission for a specified use case. It involves a pair of locations that need to be covered, considering communication requirements of payload and CNPC links.
The flying direction of UAV is usually optimized to meet the application requirements, based on some cost function involving BS locations, association sequence and mission type~\cite{challita2018deep,zhang2018cellular,senadhira2019uplink}. A UAV trajectory is optimized to minimize the UAV flight time by ensuring that the UAV is always connected to at least one BS, often with some discontinuity tolerance limit~\cite{bulut2018trajectory}. An optimization of flight path with above assumption is known as communication-aware trajectory design.

\begin{figure*}
        \begin{subfigure}[b]{0.45\textwidth}        	   	
                \includegraphics[width=8cm,height=6cm]{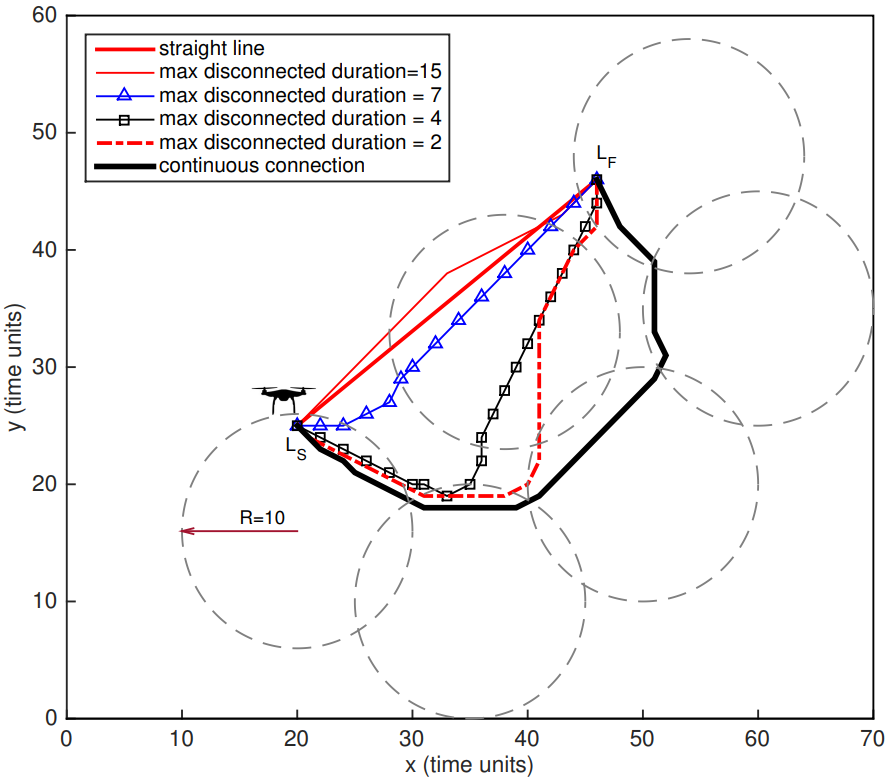} 
                \caption{UAV trajectory with first cell layout}
                \label{fig:trajresult1}
        \end{subfigure} \hfill
        \begin{subfigure}[b]{0.45\textwidth}
       	    \includegraphics[width=8cm,height=6cm]{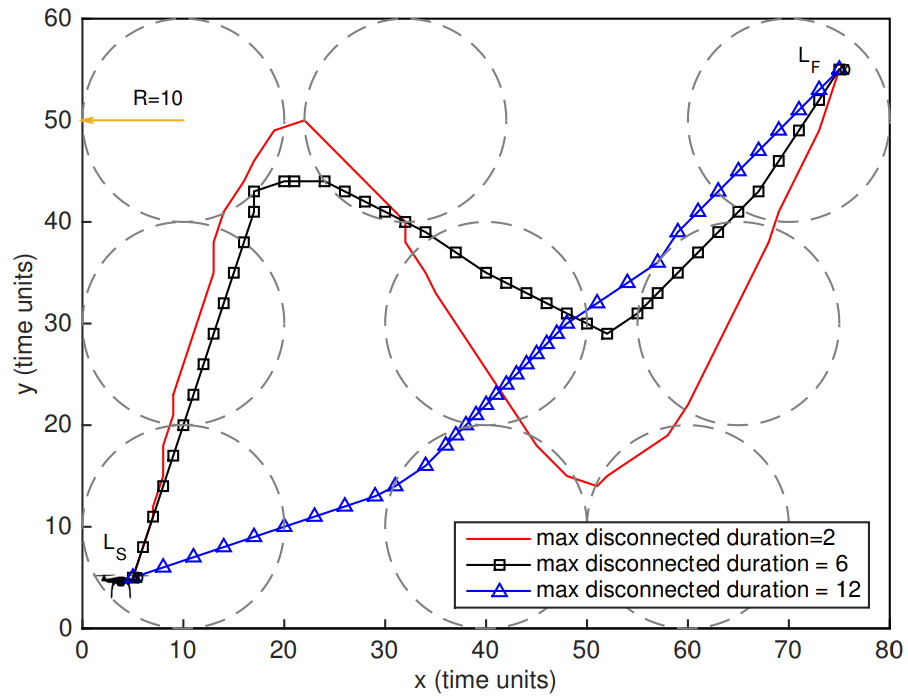}
                \caption{UAV trajectory with second cell layout}
                \label{fig:trajresult2}
        \end{subfigure} \hfill
		\captionsetup{justification=centering}
        \caption{UAV trajectory for two different cellular layouts with respect to a discontinuity threshold~\cite{bulut2018trajectory}}\label{fig:trajresults}
\end{figure*}

\begin{table*}[width=1\textwidth,cols=4,pos=h]
  \caption{Reference works on trajectory optimization for cellular-connected UAVs}
  \setlength{\tabcolsep}{8pt}
  \begin{tabular*}{\tblwidth}{@{} CCC@{} }
   \toprule
    Key Considerations & Approach Taken & Goals of Optimization \\
   \midrule
   \parbox{5cm}{\vspace{.25\baselineskip} Disconnectivity constraints~\cite{bulut2018trajectory}\vspace{.25\baselineskip} } & \parbox{6cm}{\vspace{.25\baselineskip} Dynamic Programming based approximate solution with low complexity\vspace{.25\baselineskip} } & \parbox{6cm}{\vspace{.25\baselineskip} Minimize the UAV trajectory distance without staying out of coverage for certain threshold\vspace{.25\baselineskip} } \\   \hline
   \parbox{5cm}{\vspace{.25\baselineskip} Connectivity constraints~\cite{zhang2018cellular}\vspace{.25\baselineskip} } & \parbox{6cm}{\vspace{.25\baselineskip} Graph connectivity-based approach\vspace{.25\baselineskip} } & \parbox{6cm}{\vspace{.25\baselineskip} Minimize the UAV’s mission completion time by optimizing the trajectory\vspace{.25\baselineskip} } \\ \hline
   \parbox{5cm}{\vspace{.25\baselineskip} Interference-aware~\cite{challita2018deep}\vspace{.25\baselineskip} } & 
   \parbox{6cm}{\vspace{.25\baselineskip} Deep reinforcement learning algorithm based on echo state network (ESN) cells\vspace{.25\baselineskip} } & 
   \parbox{6cm}{\vspace{.25\baselineskip} Maximize the energy efficiency, minimize the wireless transmission latency and interference on ground network, minimize the time needed to reach destination\vspace{.25\baselineskip} } \\ 
\bottomrule
  \end{tabular*}
  \label{traj}
\end{table*}

The rural and unpopulated areas with poor or no cellular connectivity impact UAV trajectory, as the persistent connection controlling the UAV might be interrupted. Additionally, UAVs operations in mmWave bands of 5G suffer from greater path loss and blockages leading to interrupted connections during mission path. Fig.~\ref{trajectory} demonstrates the need for communication-aware trajectory design in cellular-connected UAVs consisting of many ground BSs and a single UAV. Assume that the UAV has to cover a path from start position S to final position F. As shown in the figure, the coverage from all the ground stations does not fully meet the connection requirement and may suffer from discontinuity. Following are two main observations that complicate this mission path and must be accounted in the communication-aware trajectory design:

\begin{itemize}
\item The flight path may not be a linear or straight path from S to F, although it is distance-optimal. The UAV must exhibit persistent connection with cellular networks during flight path, thereby making it non-linear or curved.
\item The optimal path may pass beyond cellular coverage and hence, proper tolerance limits have to be applied before the UAV connection is interrupted. The cases, where the discontinuity duration exceeds beyond the acceptable tolerance limit, the UAV fails to accomplish the given mission being unable to maintain a successful connection to cellular network.
\end{itemize}


\subsubsection{Associated Works and Illustrative Results}
Table~\ref{traj} highlights the existing literature for UAV trajectory optimization. The authors in~\cite{bulut2018trajectory} formulate an approximate optimum trajectory finding problem for cellular-connected UAVs without exceeding a given discontinuity tolerance limit between a pair of locations. The problem is solved by a dynamic programming approach having low computational complexity and is shown to achieve close to optimal results. Fig.~\ref{fig:trajresults} demonstrates the UAV trajectory for two different cellular layouts with respect to a discontinuity threshold. It is clear that, the UAV respects this threshold limit to generate the flying coordinates for trajectory. Threshold value of zero (\emph{i.e.,} continuous connection) generates a trajectory that must pass through the cellular coverage, as shown by a dark black line in Fig.~\ref{fig:trajresult1}. When the threshold value is 15 time units (shown by a red line in Fig.~\ref{fig:trajresult1}), then the trajectory tries to minimize the distance covered and nearly follows a straight path for distance optimization. Similar justifications are also valid for second cellular layout shown in Fig.~\ref{fig:trajresult2}.


\subsection{Security Challenges}
\subsubsection{Preliminary}
Cellular-connected UAVs are usually equipped with a multitude of sensors that collect and disseminate data. This provides numerous opportunities to expose them to vulnerabilities. These flying platforms are prone to cyber physical attacks, with an intention to steal, control and misuse the UAV payload information by reprogramming it for undesired behaviour. For instance, in business use case such as goods delivery, the attacker can gain physical access to the customer package as well as to the UAV device. Existing information security measures are not well suited for cellular-connected UAVs, because these measures do not take into account possible threats imposed on numerous on-board sensors and actuator measurements of UAVs~\cite{rani2016security,choudhary2018intrusion}. An attacker can manipulate the UAV's communication and control system, thereby making it very difficult to bring it back online. Thus, it is crucial to develop new protection methodologies to avoid aforementioned intrusions and hacking procedures~\cite{challita2019machine}.


\subsubsection{Associated Works and Illustrative Results}
Inspired by the efficacy of the AI and ML-empowered approaches, in~\cite{challita2019machine}, the authors presented various security challenges focusing from the viewpoint of three different cellular-connected UAV applications. They are - (i) UAV-based delivery systems (UAV-DS), (ii) UAV-based real-time multimedia streaming (UAV-RMS) and (iii) UAV-enabled intelligent transportation systems (UAV-ITS). In order to solve this challenge, the authors proposed an artificial neural network (ANN) based solution approach which adaptively optimizes the network changes to safeguard the resource and UAV operation. 
\begin{itemize}
\item UAV-DS: These systems are vulnerable to cyber-physical attacks where the delivery of goods is compromised. The malicious intruder takes control of the UAV with an intention to destroy, steal or delay the transported goods. Even the UAVs can be physically attacked to acquire the goods being transported along with physical UAV assets.
\item UAV-RMS: UAV-enabled VR, online video transmission and online tracking are some of the use cases in this type of application. An attacker can manipulate the identity of the UAV and transmit disrupted information to the control station using their identities. In a large-scale deployment of UAVs, the control station must process the multi-media files incurring a large delay and burdening high utilization of computational resources.
\item UAV-ITS: This application ensures road safety, traffic analysis to monitor accidents, track compromised vehicles, etc. Such benefits are achieved by a swarm of cellular-connected UAVs cooperating to capture needful data during mission. An attacker can choose to send an unidentified UAV to join the swarm of UAVs to steal the information or initial self-collision to disrupt the UAV-UAV communication. Such attacks can bring serious consequences to the entire mission.
\end{itemize}

In~\cite{choudhary2018intrusion}, the authors have presented a brief survey of state-of-the-art intrusion detection system (IDS) mechanisms for networked UAVs. It highlights existing UAV-IDS approaches and areas that need attention for building a secure UAV-IDS system.



\nomenclature{U2G}{UAV-to-Ground}
\nomenclature{G2U}{Ground-to-UAV}
\nomenclature{U2U}{UAV-to-UAV}
\nomenclature{mmWave}{Millimetre Wave}
\nomenclature{VNF}{Virtual Network Function}
\nomenclature{SFC}{Service Function Chaining}
\nomenclature{RSRP}{Referenced Signal Referenced Power}
\nomenclature{RACH}{Random Access Channel}
\nomenclature{ICIC}{Inter-Cell Interference Coordination}
\nomenclature{ULA}{Uniform Linear Antenna Array}
\nomenclature{SCADA}{Supervisory Control and Data Acquisition}
\nomenclature{NF}{Network Function}
\nomenclature{MEC}{Mobile Edge Computing}
\nomenclature{IoT}{Internet of Things}
\nomenclature{QoS}{Quality of Service}
\nomenclature{RSSI}{Received Signal Strength Indicator}

\subsection{Summary of Lessons Learnt}
In this Section, we have seen that, despite of the benefits and wide popularity of cellular-connected UAVs, there are several challenges and operational complications that needs to be investigated to realize their true potential. The important lessons learnt from this Section are listed as follows:
\begin{itemize}
\item The varying altitude of UAV necessitates a 3D wireless coverage model for base stations, because the current design of terrestrial base station is highly optimized for ground users. Typically, the UAVs fly higher than base station creating LoS links that are prone to be interfered from other neighbouring base stations. Proper interference management becomes challenging and critical in terms of harmonious coexistence between aerial UEs and ground UEs simultaneously.
\item UAVs are highly mobile and mainly served by the side lobes of existing base stations. This produces a peculiar cell association and increased handover rates, completely different than that of ground users. The mobility of UAV in 3D space necessitates enhanced cell selection and seamless handover patterns to optimize its operation.
\item The battery life of a UAV is limited. During the mission, UAV must intelligently plans its trajectory from initial to destination location considering application and use case. The key performance metrics, such as maximum allowed time to complete the mission, persistent cellular connectivity, QoS guarantees, energy consumption, etc. are some of the factors the UAV must respect during its mission. Hence, trajectory optimization is an essential aspect of UAV mission.
\item While carrying out sensitive and real-time critical tasks, UAVs are prone to security threats and cyber physical attacks. Any malicious attempt to steal, misuse or control the UAV, can trigger undesirable situations and cause loss of confidential and private assets. Such security threats require stringent protection measures, guidelines and regulations by the operator.
\item Before 5G/B5G cellular systems can really benefit from the UAV technology, above mentioned technical integration challenges demand a thorough investigation and practical solutions.
\end{itemize}

\section{Synergies of 5G/B5G innovations for Cellular-connected UAVs}
By design, cellular-connected UAVs are expected to be controlled and managed remotely by a Ground Control Station (GCS). Depending upon the UAV application and use case, the UAVs carry out different missions, which require unique networking characteristics. In general, these networking requirements are very tightly coupled with the use case and hardware infrastructure support. Especially, multi-UAV systems comprising many functional and coordinated UAVs, establishing the reliable and secure communication path as well as the design and development of efficient reconfigurable network architectures is a challenging issue. 

The key innovations of 5G/B5G systems are the cloudification and virtualization of network resources through Software Defined Networking (SDN), Network Function Virtualization (NFV), Service Function Chaining (SFC), network slicing, and physical layer improvements. SDN segregates the control functions and forwarding functions of a device. It allows softwarization of the control functions, thereby making the network programmable. NFV transforms the traditional network services into software based solutions (Virtual Network Functions i.e., VNFs) that can be dynamically deployed on a general purpose hardware platforms. SFC is a chain of simple and smaller network functions that must follow an execution sequence to realize a complex and large network function. Future 5G-centric networking applications and services are driven by programmable network architectures, where softwarization and cloudification of network functions are the key enablers. Therefore, the above-mentioned 5G innovations are envisioned as a part of the cellular-connected UAV applications and will be detailed in this Section. Specifically, Section~\ref{netarch} focuses on the envisioned network architectures for cellular-connected UAVs. The hardware and physical layer improvements are discussed in Section~\ref{phy5g}.

\begin{table*}[width=1\textwidth,cols=3,pos=h]
\caption{Envisioned Network architectures of Cellular-connected UAV}
\begin{tabular*}{\tblwidth}{@{} CCC@{} }
\toprule
Technology & Principle & Benefits \\
\midrule
\parbox{3cm}{\centering NFV-Oriented \\ ~\cite{nogales2018nfv},~\cite{nogales2018adaptable},~\cite{czentye2018controlling}} & \parbox{5cm}{Decouples the hardware and software that exists in traditional vendor network setting} & \parbox{8cm}{
\begin{itemize}
\item Greater flexibility for NF deployment 
\item Dynamic service provisioning
\item Easily deployable and well scalable 
\item Efficient allocation to general purpose hardware
\end{itemize}} \\ \hline
\parbox{3cm}{\centering MEC-Oriented \\ ~\cite{zhou2020mobile},~\cite{cao2018mobile},~\cite{grasso2019fleet}} & 
\parbox{5cm}{The cloud computing capabilities are placed close to edge of mobile network} & \parbox{8cm}{
\begin{itemize}
\item Significant reduction in data exchange cost
\item Computational offloading to local servers
\item Improvement of QoE for end users
\end{itemize}}  \\ \hline
\parbox{3cm}{\centering IoT-Oriented \\ ~\cite{motlagh2016low},~\cite{ejaz2019unmanned},~\cite{motlagh2017uav},~\cite{lagkas2018uav}} & \parbox{5cm}{Connecting massive number of diverse, smart devices to 5G/B5G cellular network}
& \parbox{8cm}{
\begin{itemize}
\item Assists efficient decision making on huge data 
\item Extracting meaningful information for end users
\item Control automation without human intervention
\item Information sharing and communication 
\end{itemize}} \\ \hline
\parbox{3cm}{\centering Service-Oriented\\ ~\cite{royo2008service},~\cite{koubaa2019dronemap},~\cite{besada2019drones}} &  \parbox{5cm}{Network services are provided over a communication protocol that is independent of vendors or product} & \parbox{8cm}{
\begin{itemize}
\item Improves the modularity of application
\item Transforms monolithic networking application into a set of microservices
\item Each microservice is a basic unit of functionality
\end{itemize}}  \\ 
\bottomrule
\end{tabular*}
\label{relnetarch}
\end{table*}

\subsection{Network Architectures} \label{netarch}
Based on the key enablers of future 5G-centric networking applications, the cellular-connected UAV network architectures can be summarized in the following groups. They are (i) NFV Oriented, (ii) MEC Oriented, (iii) IoT Oriented, (iv) Service Oriented (SOA). In next subsection, we highlight the respective architectures and existing works. Table~\ref{relnetarch} shows the glossary of related works for each of these network architectures. 


\begin{figure*}[ht] 
\minipage{0.48\textwidth}
\centering
\includegraphics[width=7.5cm,height=5.5cm]{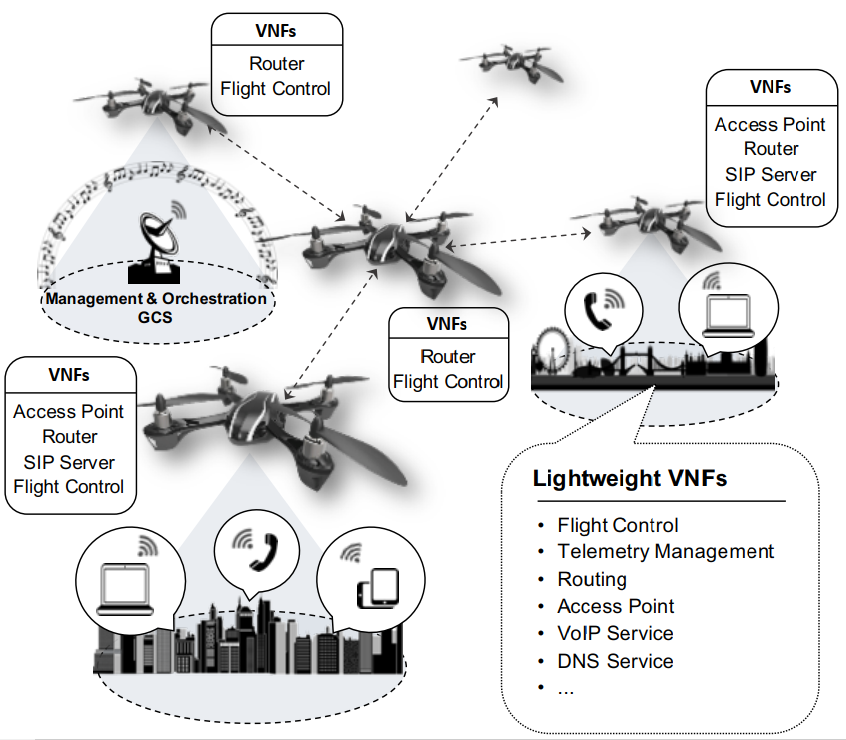}
\caption{NFV based achitecture for UAVs~\cite{nogales2018nfv}}
\label{arch2label}
\endminipage\hfill
~
\minipage{0.48\textwidth}
\centering
\includegraphics[width=7.5cm,height=5.5cm]{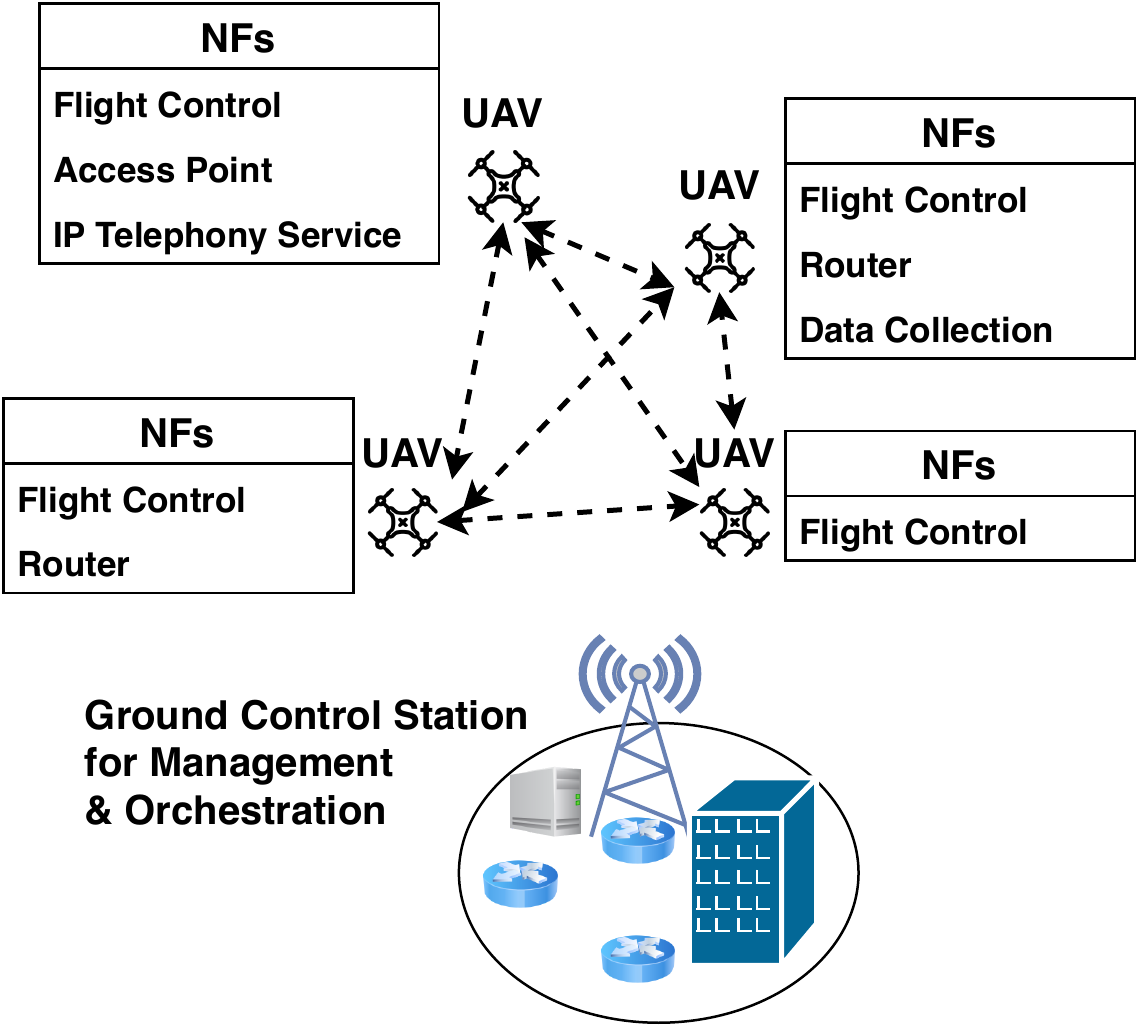}
\caption{UAVs with diverse network functions}
\label{arch1label}
\endminipage\hfill
\end{figure*}

\begin{figure*}[t]
\centering
\includegraphics[width=0.85\linewidth]{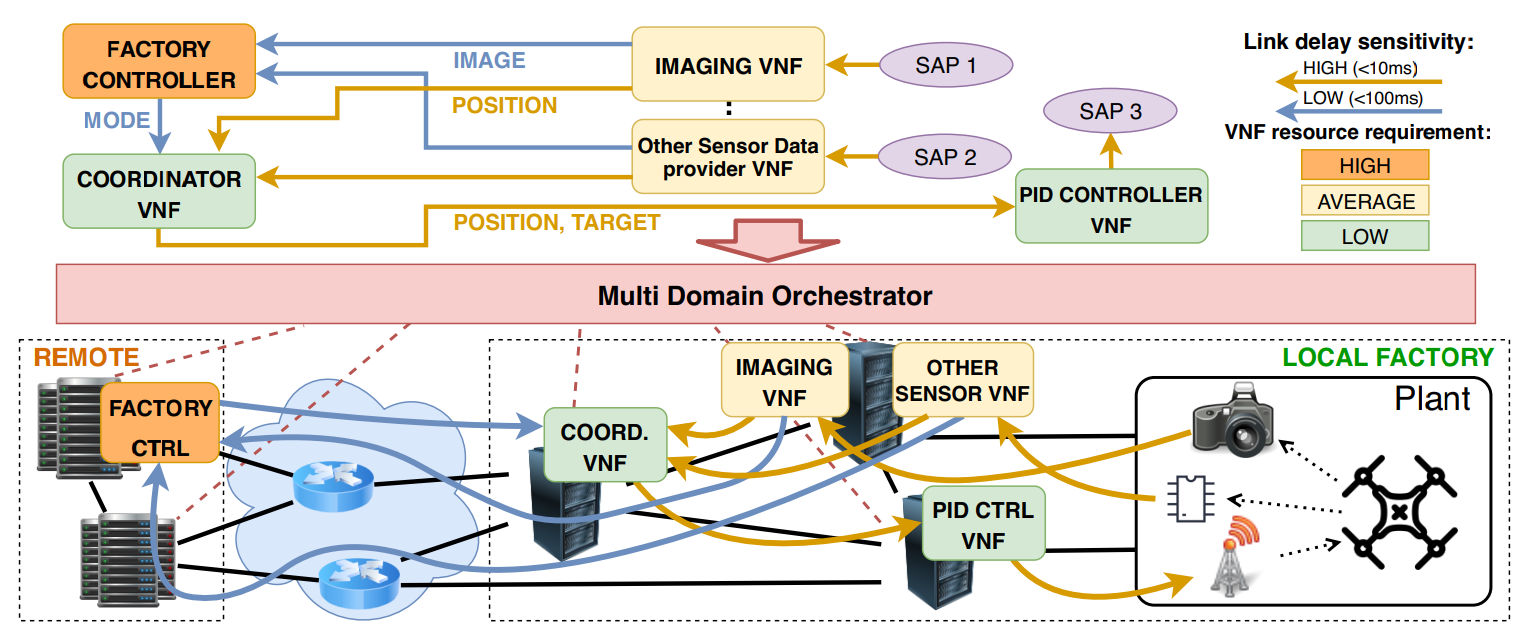}
\caption{Deployment on multi-domain UAV services~\cite{czentye2018controlling}}
\label{arch3label}
\end{figure*}

\subsubsection{NFV Oriented Architectures}
In~\cite{nogales2018nfv}, the authors present the feasibility of an agile, automated and cost-effective UAV deployment architecture carrying out heterogeneous missions with the help of NFV technology. This work proposes an adaptable way to achieve a reconfigurable UAV management system, which is capable of carrying out missions with varying objectives. For example, some UAVs could incorporate a VNF that provides access point connectivity services, another VNF for network layer routing functionalities, a third VNF for flight control system that can be easily upgraded as per the changing needs of the mission. The work is validated by a prototype built upon open-source technologies. The high-level architecture of such a system is shown in Fig.~\ref{arch2label}.
As shown in the figure, the communication infrastructure formed by a set of UAVs, where the mission planner used a MANO NFV framework (defined by ETSI), installed at ground station to flexibly deploy a set of VNFs over the set of UAVs. Overall design of such a system consists of the following components:
\begin{itemize}
	\item Management and Orchestration (MANO):- It is located with the GCS and realized by Open Source MANO (OSM) Release TWO. It contains all the necessary functionalities of service orchestration and VNF manager as per ETSI NFV reference architecture~\cite{etsi2014002}. 
	OpenStack Ocata is used for VIM. Both OSM and VIM were deployed in mini-ITX computer having 4 Gb Ethernet  ports, 8 GB RAM, Intel Core i7 2.3 GHz, 128 GB SSD with DPDK support. 
	\item UAV hardware and software:- It provides the infrastructure support for execution and deployment on light-weight VNFs. It is realized by Parrot AR.Drone 2.0 carrying single board Raspberry Pi 3 Model B.
	\item Mission Planner:- It is located at the GCS and defines the nature and characteristics of different network services or network functions (NFs) to be deployed along with their placement policies. It also interfaces with MANO component to call for the light-weight VNF deployment on set of UAVs.
\end{itemize}

In order to carry out routing of VNFs to different target UAVs, LXC Linux containers on Ubuntu OS are used. Each routing VNF requires resources of 1 vCPU, 128 MB RAM and 4 GB storage. 

The authors in~\cite{nogales2018adaptable} have presented a practical NFV based approach to support UAV multi-purpose deployment, which can be rapidly configured according to the need of the civilian mission. They have considered the UAVs to provide infrastructure and hardware that enable agile integration of network functions at deployment time by a network operator. As shown in Fig.~\ref{arch1label}, a set of UAVs could be used for providing communication infrastructure (virtual access points) in case of disaster or can be used in SAR operation in a remote area. The mission specific UAV behaviours are softwarized as network functions and installed to UAV infrastructure (hardware) at the time of deployment. Some network functions pertaining to mandatory features of any UAV such as flight control and telemetry are installed on all UAV hardware, irrespective of the mission. The implementation of the system prototype and the light-weight VNF is done using open-source software technologies. The orchestration and life-cycle management of light-weight VNF is done by OSM Release FOUR. OpenStack Ocata version is used for realizing the virtual infrastructure layer (VIM). The virtual machine environment runs mini-ITX computer which consists of Intel Core i7 2.3 GHz processor, 4Gb Ethernet ports, 16 GB RAM, 128 GB SSD. The UAV hardware platform consists of DJI Phantom 3 carrying a Raspberry Pi 3 Model B computing board and serves the platform for execution of light-weight VNFs needed for specific mission. 

A software based service architecture running on a distributed cloud environment is demonstrated in~\cite{czentye2018controlling}. In this demonstration, an Industry 4.0 application controlling the indoor drones is considered for study. The application is implemented using SFC orchestrated by a multi-domain orchestrator known as ESCAPE. The orchestrator is able to setup and configure VNFs onto the physical UAV boards according to mission's policies and requirements. The proposed implementation is shown in Fig.~\ref{arch3label}.
The deployment occurs when the service requests are triggered to ESCAPE as per requirement. OpenStack is used for running the cloud environment and few laptop hosts are used as edge execution machines by Docker platform. High level commands such as take-off, land, fly are used for controlling the UAV behaviour from factory controller. 

\begin{figure*}[htb!]
        \begin{subfigure}[b]{0.32\textwidth}        	   	
                \includegraphics[width=5.7cm,height=4.7cm]{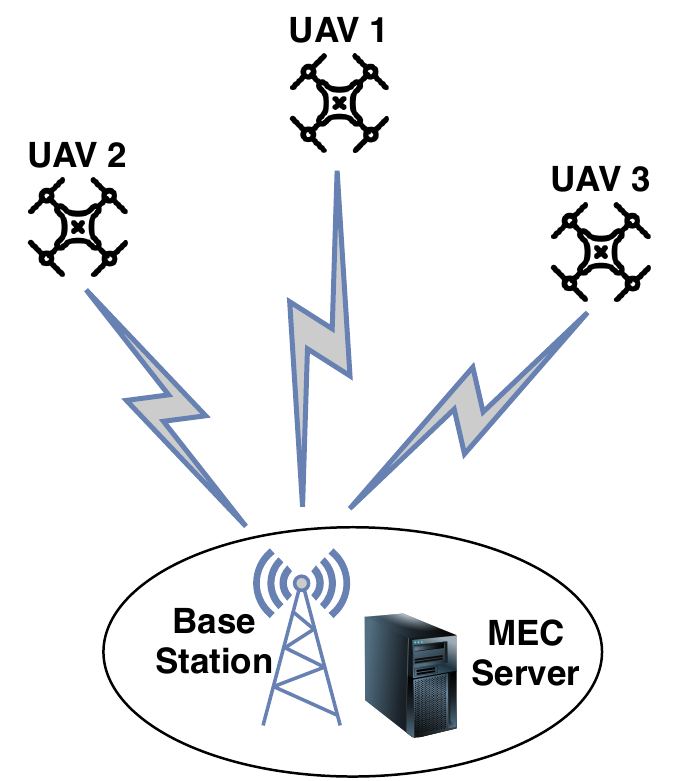}
                \caption{MEC based Architecture~\cite{zhou2020mobile}}
                \label{fig:mecarch1}
        \end{subfigure} \hfill
        \begin{subfigure}[b]{0.32\textwidth}    
                \includegraphics[width=5.7cm,height=4.7cm]{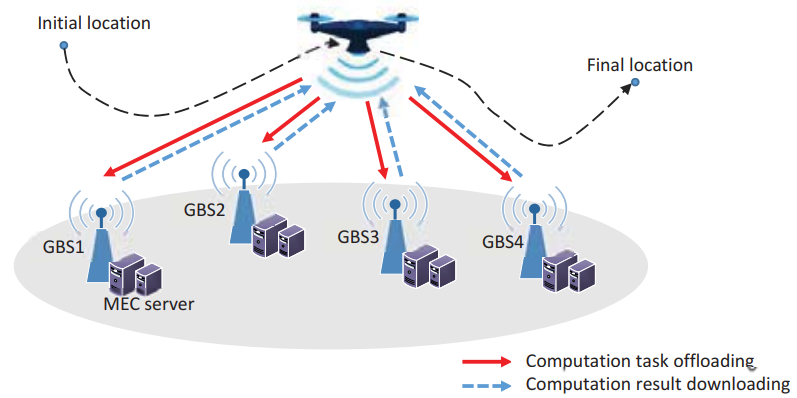}
                \caption{UAV Computational Offloading~\cite{cao2018mobile}}
                \label{fig:mecarch2}
        \end{subfigure}  \hfill
        \begin{subfigure}[b]{0.32\textwidth}
                \includegraphics[width=5.7cm,height=4.7cm]{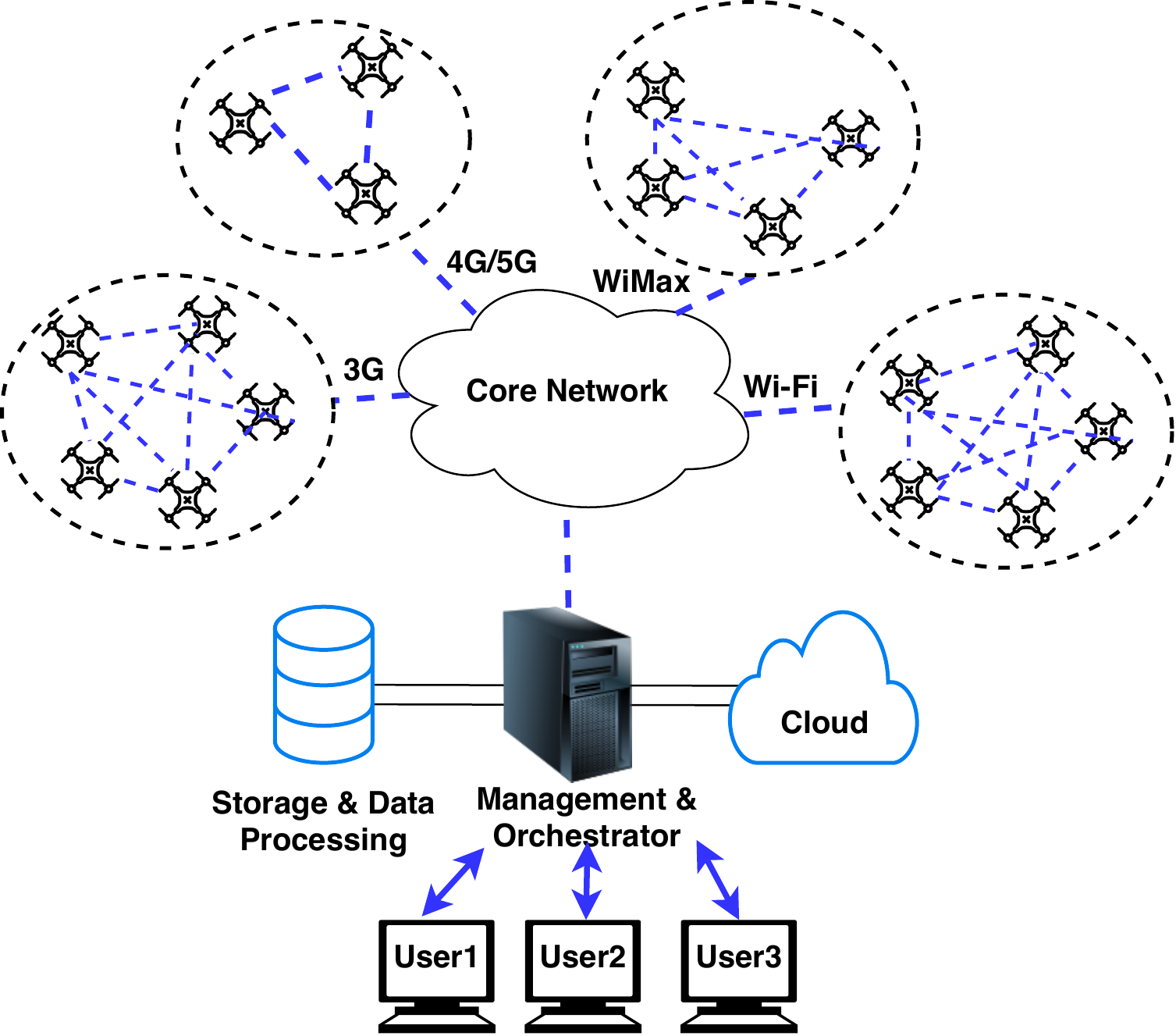}
                \caption{UAV-based IoT Platform}
                \label{fig:uaviot1}
        \end{subfigure} \hfill
        \caption{Network Architectures of Cellular-connected UAV}\label{fig:netarchs5g}
\end{figure*}


%

\subsubsection{MEC Oriented Architectures}
In general, UAVs possess physical constraints in terms of computational capability, storage and battery capacity. MEC has been identified as one of the promising techniques to deal with the limitations of low computational capability and restricted battery capacity of flying UAV. Some examples of resource-intensive tasks are trajectory optimization, object recognition, AI processing in crowd-sensing. Due to the limited onboard resources of the UAVs, computation of above resource intensive tasks are not very efficient. Hence, in such case, edge-cloud based network architectures provide substantial improvements for operations of cellular-connected UAVs.\\

In~\cite{zhou2020mobile}, the authors presented a UAV-enabled MEC architecture applicable for cellular-connected UAVs. Fig.~\ref{fig:mecarch1} illustrates this architecture, where the UAV has some computational task to be executed. This task can be offloaded to the MEC server located with the ground station and, after the computation, obtained results can be sent back to UAV for their exploitation. Depending upon the volume of the offload, there can be two modes of operation: (i) partial mode, and (ii) binary mode. In partial offload mode, the whole task is split into two parts. One part is executed locally and the other part is executed by the MEC server (e.g., face recognition use case). In binary offload mode, each task is executed as one unit, irrespective of whether it is done locally or at the MEC server (e.g., channel state information (CSI) estimation). Both of these offload modes have advantages and drawbacks. The selection of the suitable mode depends on the nature of computational task being performed, UAV structure and characteristics. 

Considering the use case of trajectory optimization and computational offloading in cellular-connected UAV, the work in~\cite{cao2018mobile} presents a novel MEC setup, where the UAV needs to offload some of its processing task to the ground station. The UAV flies from an initial location to a destination location and offload the task to selected ground base stations during the trajectory. The goal of the MEC setup is to minimize the total time for UAV mission considering the maximum speed and ground station capacity constraints. This setup is shown in Fig.~\ref{fig:mecarch2}.

In reference work~\cite{grasso2019fleet}, the authors proposed a 5G network slicing concept extend to video monitoring with UAVs having MEC facilities. The surveillance area is divided into multiple zones and a set of UAVs are assigned the task to monitor a specific zone. The MEC enabled UAVs could offload the captured data and video streams with acceptable quality and performance. 

%


\subsubsection{IoT Oriented Architectures}
In~\cite{motlagh2016low}, the authors envision a heterogeneous UAV network architecture, where UAVs are used to deliver value-added IoT services from the sky. The UAVs are considered as key enabler of IoT framework that are deployed by following a specific vision. Each UAV is equipped with various IoT sensors or camera to gather data. The deployment spans across a large area, where UAVs are grouped to form UAV clusters (because of close geographical proximity or mission type or altitude). A fixed UAV is designated as cluster head (CH), and is mainly responsible for disseminating collected data to the other UAVs or orchestrator via core network. The core network performs the intelligent decisions and employs algorithms for efficient processing the data gathered from UAV sensors. The high level architecture schematic of this proposal is shown in Fig.~\ref{fig:uaviot1}.

In~\cite{ejaz2019unmanned}, the authors presented the network architecture of a UAV-enabled IoT framework developed for disaster mitigation. In this case, the UAV not only acts as a flying base station in emergency situation, but also behaves as a cellular-connected UAV for information dissemination in scenarios such as wildfire or environmental losses. The framework consists of three main components, (i) ground-IoT network, (ii) connectivity of UAV and ground-IoT network and (iii) data analytics. 

The authors in~\cite{motlagh2017uav} demonstrated a UAV-based IoT framework for crowd surveillance application which collects the data and performs facial recognition to track and identify suspicious activities in a crowd. The fleet of UAVs are managed by a centralized orchestrator component.


\subsubsection{Service Oriented Architectures (SOA)}
The work in~\cite{royo2008service} demonstrates the design and development of a UAS Service Abstraction Layer (USAL) for UAV which implements different types of missions with minimal re-configuration time. USAL contains a set of predefined useful services that can be configured quickly according to the requirements of civil mission. The architecture is service oriented, and the service abstraction layer provides the re-usability of the system. The mission functionalities are split into smaller parts and are implemented as independent services. USAL replies on a middleware that manages the services and their communication needs. USAL may contain a large number of services, however, all of them need not be present. Depending upon the mission, suitable services can be loaded and activated to meet the objective of mission. \\

\begin{figure}[pos=b]
\centering
\includegraphics[width=1\linewidth]{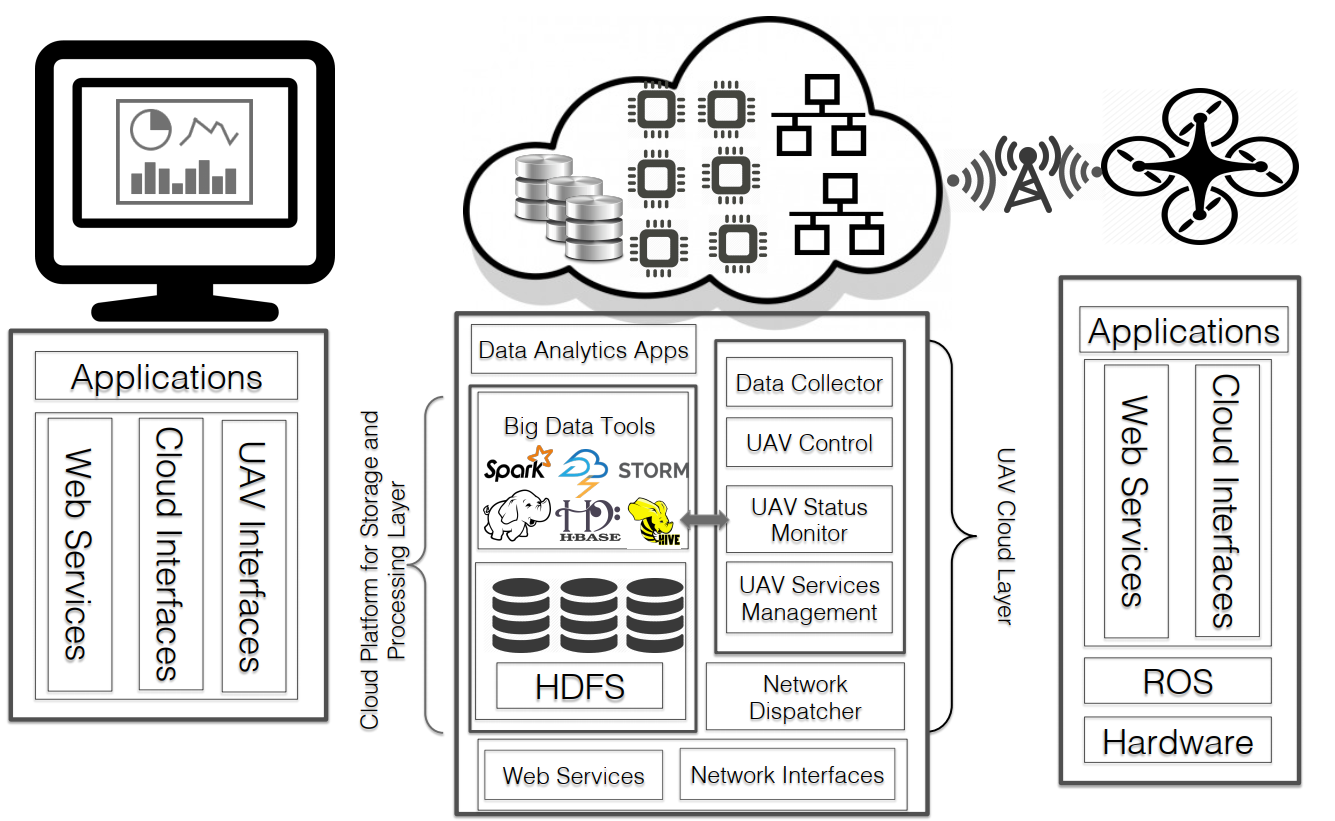}
\caption{Service Oriented System Architecture of UAV~\cite{koubaa2019dronemap}}
\label{soa}
\end{figure}

In~\cite{koubaa2019dronemap}, the authors presented Dronemap Planner, a service-oriented cloud based UAV management system, which performs overall management of UAVs over Internet and control their communication and mission. It virtualizes the access mechanism of UAVs via REST API or SOAP. It uses two communication protocols: (i) MAVLink and (ii) ROSLink. The objective of designing such a system is to provide seamless control to monitor UAVs, offload compute intensive tasks to cloud platform, and dynamically schedule the mission on demand. The cloud computing model creates an elastic model that scales well with the numbers of UAVs as well as with the offered services. Fig~\ref{soa} shows the schematic of system architecture developed in this study.

\subsection{Hardware and Physical layer consideration} \label{phy5g}
The performance of cellular-connected UAVs in 5G networks significantly depends on the underlying physical layer signal processing. In this section, we highlight the candidate physical layer techniques that influence the UAV communication. The key techniques are massive MIMO (Multiple Input Multiple Output) antenna, mmWave communication (3-300 GHz), beamforming and beam division multiple access (BDMA), as well as some new modulation schemes. 
In 4G LTE, Orthogonal frequency division multiplexing (OFDM) and Orthogonal frequency division multiple access (OFDMA) are predominantly used for multiplexing and multiple access method. 5G/B5G networks is considering new waveforms to support efficient air interface~\cite{mitra20155g}. These new waveforms are superior than OFDM and no longer require strict orthogonality and synchronization. Table~\ref{modscheme} provides a brief categorization of different waveforms for 5G from implementation perspective. 

\begin{table}
  \caption{Candidate waveforms for 5G}
  \begin{tabular*}{\tblwidth}{@{} LL@{} }
   \toprule
    Scheme & Short Description\\
   \midrule
\parbox{2.5cm}{\vspace{.25\baselineskip}Generalized Frequency Division Multiplexing (GFDM)\vspace{.25\baselineskip}} & \parbox{5.3cm}{\vspace{.25\baselineskip}It is a block-based modulation approach where the available bandwidth is either divided into several narrow bandwidth subcarrier or few subcarriers with high bandwidth for each.\vspace{.25\baselineskip}}   \\   \midrule
\parbox{2.5cm}{\vspace{.25\baselineskip}Universal Filter Bank Multi-carrier (UFMC)\vspace{.25\baselineskip}} & \parbox{5.3cm}{\vspace{.25\baselineskip}Multicarrier signal format to handle loss of orthogonality at receiver end. It uses sub-band short duration filters.\vspace{.25\baselineskip}} \\   \midrule
\parbox{2.5cm}{\vspace{.25\baselineskip}Filter Bank Multicarrier (FBMC)\vspace{.25\baselineskip}} & \parbox{5.3cm}{\vspace{.25\baselineskip}It uses a preamble burst based approach to ensure flexible resource allocation.\vspace{.25\baselineskip}}   \\ \midrule
\parbox{2.5cm}{\vspace{.25\baselineskip}Biorthogonal Frequency Division Multiplexing (BFDM)\vspace{.25\baselineskip}} &  \parbox{5.3cm}{\vspace{.25\baselineskip}It uses a relaxed form of orthogonality where transmitter and receiver are bi-orthogonal. In other words, the transmitted and received pulses have to be pairwise orthogonal. BFDM is more robust than OFDM.\vspace{.25\baselineskip}} \\
   \bottomrule
  \end{tabular*}
  \label{modscheme}
\end{table}

\subsubsection{5G NR}
5G new radio (NR) is a new radio interface and radio access network which is designed and developed for advanced cellular connectivity. It utilizes novel modulation schemes and access technologies that help the underlying system to cater to high data rate services and low latency requirements. The first version of 5G NR started in 3GPP Rel-15. 5G NR supports the frequency ranges in sub 6 GHz or in mmWave range (24.25 to 52.6 GHz). It has greater coverage and enhanced efficiency because of beamformed controls, MIMO and access mechanisms. 5G NR is expected to cater to three broad categories of services i.e., (i) extreme mobile broadband (eMBB), (ii) ultra-reliable low latency communication (URLLC) and (iii) massive machine type communication (mMTC).  
Specifically, the expectations for each of the mentioned scenarios are:
\begin{itemize}
	\item For eMBB use case scenario, the data rate is promised as 100 Mbps and three time more spectral efficiency than 4G systems. It will be able to support a device that moved with a maximum speed of 500 km/h. 
	\item For URLLC use case scenario, the goal is to achieve 1 ms latency with reliability 99.999\%. It means, the reliability of the wireless link will not be met, if more than one data unit out of $10^5$ data units does not get delivered within 1 ms.
	\item For mMTC use case scenario, the density of devices that 5G will be able to handle will reach nearly 1000000 per square kilometer. 
\end{itemize}

URLLC ensures strict latency and reliability requirements for the application. 5G NR focuses on framing, packetization, channel coding and diversity enhancements for achieving URLLC. One of the most vital scenarios of URLLC is the remote piloting of cellular-connected UAVs in BVLoS range. Package delivery, remote surveillance and border patrolling are some of the use cases that demand UAV operations in BVLoS range. Due to the changing altitude, velocity and distance between remote UAV and ground station, the URLLC requirements may vary. URLLC is the key use case scenario to enable BVLoS UAV operations,
which assist in safe UAV piloting to avoid crashes, obstacles etc. Cellular-connected UAVs can benefit from 5G NR design as it offers dominant uplink data transmissions from UAV to ground BS, especially for many demanding use cases pertaining to streaming, surveillance, imaging etc. The downlink data transmission requirement is much smaller in contrast to uplink. Moreover, the sub 6 GHz and millimetre wave spectrum could potentially be used for the downlink and uplink respectively, considering the asymmetric traffic requirements.

\subsubsection{Massive MIMO}
Massive MIMO is a promising technology that consists of a large number of controllable antenna arrays. It is supported by 3GPP in Rel-15 for 5G NR. 5G will exploit full benefits of MIMO by leveraging the uncorrelated and distributed spatial location of cellular-connected UAVs, as well as ground users. Massive MIMO enhances the signal strength, where multiple data streams can include unique phase and weights to the waveforms to be constructively generated at the UAV receiver~\cite{garcia2019essential,geraci2018understanding}. It minimizes the interference to other cellular-connected UAV receivers. \\

The work~\cite{garcia2019essential} presents an evaluation of a massive MIMO system for cellular-connected UAVs. It demonstrates that, massive MIMO assists in harmonious coexistence of cellular-connected UAVs with ground users, supports large uplink data rates and results in consistent CNPC link behaviour. The test uses 20 MHz bandwidth in sub-6GHz licensed spectrum operating in TDD mode. Massive MIMO-enabled systems are useful to restrict the impact of interference to the existing terrestrial users. Such system requires frequent and accurate CSI updates.

\subsubsection{Millimeter-wave communications}
Millimeter-wave (mmWave) spectrum has been extensively investigated in UAV cellular communication that offers high bandwidth services using frequency spectrum above 28 GHz. The channel between cellular-connected UAVs and ground BS is typical LoS dominant and mmWave having high bandwidth are favourable for communications. However, the mmWave signals are affected by any kind of blockage, which poses several implementation challenges. Therefore, efficient beamforming and tracking are needed for cellular-connected UAV operation. \\

The work~\cite{xia2019millimeter} presents a simulated study to showcase the feasibility of using 28 GHz 5G link for public safety use case. The results claim that, it is feasible to achieve 1 Gbps throughput with sub ms latency using mmWave links when the grounds base station is situated close to the mission area. In~\cite{khawaja2017uav}, the authors conducted an analysis on the air-to-ground channel propagation for two different mmWave bands at 28 GHz and 60 GHz using ray tracing simulations. During experiment, the UAV speed was kept at 15 m/s and limited to a flight distance of 2 km. A total of four scenarios are validated such as urban, sub-urban, rural and over-sea. It is observed that, received signal strength (RSS) follows the two ray propagation model as per UAV flight path at higher altitudes. This two-ray propagation model is impacted in urban scenario due to high rise scattering obstacles. 

\subsubsection{Beamforming and BDMA}
Beamforming is a technique by which a beam (signal element directed to the users) is transmitted from the ground base station and directed to a specific user to minimize interference to other neighbouring users and maximizes the useful signal for the given user. In 5G NR, the antennas can create and exploit beam patterns for the specific cellular-connected UAV. This is of great importance, because of aerial mobility of UAVs and high LoS channel conditions from ground BSs. Beam division multiple access (BDMA) is capable to handle large number of users and to enhance the communication system capacity. In this case, a separate beam is allocated to each user. This access technology is dependent on the user positioning, location and speed of user movement.\\

Beam forming requires the base station to have more than one transceiver RF chain and the user (both aerial and ground) to have single RF transceiver. In order to support a greater number of users, the beam should be split. The key challenge is to find a way to group users that are served by a single beam without causing interference to other users at the same time. Strategies like angle of departure (AoD) or angle of arrival (AoA) help to measure the steering angle from BS to mitigate interference to some extent.\\

\begin{figure}[t]
\centering
\includegraphics[width=1\linewidth]{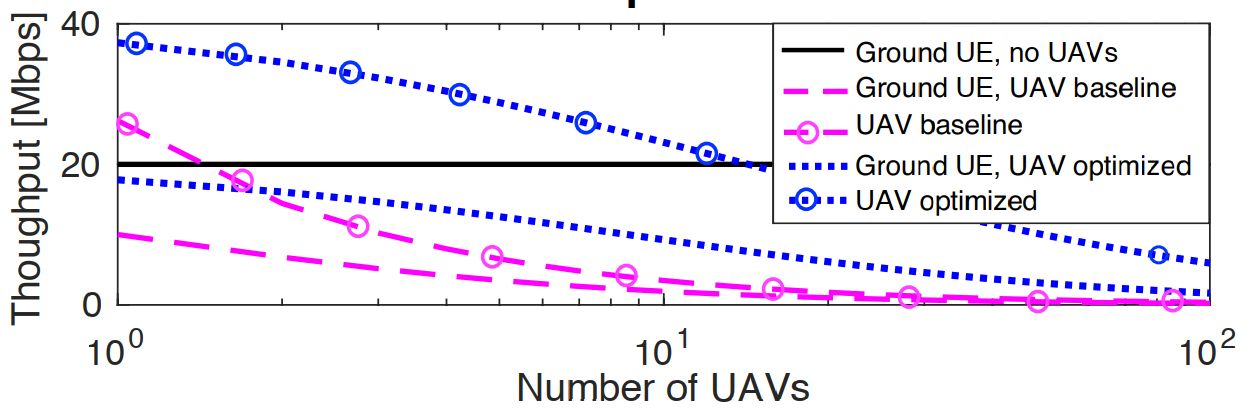}
\caption{Coexistence performance of aerial UE and ground UE, 700 MHz in rural setting~\cite{bertizzolo2020live}}
\label{beamlabel}
\end{figure}

In~\cite{bertizzolo2020live}, the authors presented a study on using steerable  directional transmitters on UAVs to evaluate the co-existence of cellular-connected UAVs with ground user. This work jointly optimized the flight path and antenna steering angle to improvise uplink throughput while minimizing interference to other neighbour base stations. The proposal is validated by a testbed setup involving $2\times2$ MIMO where wide beam transmitters are employed half-power beams at $60 \si{\degree}$ on azimuth and elevation planes and $6$ dBi forward gain. Fig.~\ref{beamlabel} shows the performance variation with respect to the throughput in presence of both ground users and aerial users. The results show that such techniques are of utmost importance when the ground and aerial users coexist at such scale. 

\subsubsection{NOMA} 
Non-orthogonal multiple access (NOMA) is a promising candidate technology for 5G wireless communication, as it leads to higher spectrum utilization than orthogonal multiple access methods (OMA). NOMA has been widely explored in UAV-assisted wireless communications, where UAV is deployed as a flying BS to serve the ground users~\cite{sohail2018non}. Few studies~\cite{mei2019uplink, rahmati2019energy} also investigated the applicability of NOMA in cellular-connected UAV network.\\

OFDMA and single-carrier (SC)-FDMA are conventional orthogonal multiple access methods (OMA) adopted as a natural choice of 4G LTE/LTE-Advanced wireless systems. The basic principles of OFDMA is to transmit the different user signals over different frequency resources, not to produce mutual interference among users. Cellular-connected UAVs coexisting with ground users benefit from such orthogonal multiple access methods, because the UAVs within a given coverage can avoid any interference to ground users by transmitting in those resource blocks that are not assigned to any ground users. Thus, the resource blocks can be allocated within the coverage area in a non overlapping manner. However, increased user density and the frequency reuse result in poor spectrum performance from such OMA methods, due to resource block scarcity. On this advent, NOMA methods allows the cellular-connected UAVs to reuse the resource blocks. In other words, NOMA is capable to serve many users at the same time/frequency resources.

NOMA employs two techniques for multiple access:
\begin{itemize}
	\item Power domain: Multiple access is based on different power levels. 
	\item Code domain: Multiple access is based on different codes.
\end{itemize}

NOMA with interference cancellation (IC) is an appealing solution to the cellular-connected UAVs because the UAVs can reuse the resource blocks that are allocated to ground users. Moreover, at high altitude, UAVs experience stronger LoS channel condition than ground users, so that BS can use IC to decode strong signal from UAVs, then subtract it to decode ground user signal~\cite{senadhira2019uplink}. 

\nomenclature{MANO}{Management and Orchestration}
\nomenclature{ETSI}{European Telecommunications Standards Institute}
\nomenclature{NOMA}{Non-Orthogonal Multiple Access}
\nomenclature{BDMA}{Beam Division Multiple Access}
\nomenclature{OMA}{Orthogonal Multiple Access}
\nomenclature{SC-OFDMA}{Single Carrier OFDMA}
\nomenclature{SINR}{Signal-to-Interference and Noise Ratio}
\nomenclature{RSRQ}{Reference Signal Reference Quality}
\nomenclature{PRB}{Physical Resource Block}
\nomenclature{RLF}{Radio Link Failure}
\nomenclature{GCS}{Ground Control Station}
\nomenclature{GSM}{Global System for Mobile Communications}
\nomenclature{GPRS}{General Packet Radio Service}
\nomenclature{COTS}{Commercial off-the-shelf}
\nomenclature{QAM}{Quadrature Amplitude Modulation}
\nomenclature{PCI}{Physical Cell Identity}
\nomenclature{EARFCN}{Absolute Radio Frequency Channel Number}
\nomenclature{UTM}{UAV Traffic Management}
\nomenclature{KPI}{Key Performance Indicators}

\subsection{Summary of Lessons Learnt}
The important lessons learnt from this section are listed as follows:
\begin{itemize}
\item Classical cellular network infrastructures are not well scalable for the diverse and growing use cases of cellular-connected UAVs. The improvements of 5G/B5G cellular networks present many candidate innovative technologies and PHY layer improvements that complements to efficient UAV operation in 5G spectrum. 
\item Based on the principles of softwarization and cloudification of networking resources, the network architectures involving cellular-connected UAV solve several practical limitations with respect to performance and scalability issue. 
\item The notable 5G advancements and trends in deploying NFV, MEC, SOA, IoT driven network architectures helps UAV technology to establish a reliable and safe communication link between ground-UAV or UAV-UAV during mission.
\item 5G-and-beyond hardware (NR) and software upgrades by cellular network operators along with the technical advancements by UAV manufacturers suitably caters to application specific latency, rate and reliability demands arising from the use cases, thereby improving overall performance of applications using cellular-connected UAVs.
\item The physical layer enhancements further supplement to the effectiveness of applications encompassing cellular-connected UAVs.
\end{itemize}


\section{Design Trials and Prototyping} \label{proto}
Experimental assessment and prototyping are time consuming and relatively complex, because they must take into consideration the deep technical aspects of any realistic deployment. There are several ongoing efforts from industry and academia that focus on experimental frameworks for cellular-connected UAVs. These efforts provide more practical insights about the underlying behaviour and complexities involved in integration of UAVs into cellular networks. Field trials and measurement campaigns  
are a cost-effective and powerful step towards the prototyping, as they help investigating the solutions to potential research problems. In the following subsections, we shed some light on these efforts and classify them into two broad groups such as (i) experimental testbeds and (ii) field trials.

\subsection{Experimental Testbeds} 
There is hardly any complete real-world testbed that fully characterizes the challenges and benefits of cellular-connected UAVs. The literature in this regard is scarce. However, several ongoing efforts are being actively pursued by researchers from both industry and academia to advance the working prototypes. It is worth mentioning that, realization of working prototypes of cellular-connected UAV mainly differ with respect to (i) the main objective for which they are built, and (ii) the features being implemented, which are also dependent on the main objective. For example, one prototype may completely focus its prototype development on investigating 5G/B5G network support to efficient UAV operations. Another prototype may prioritize its development on achieving a fail-safe, reliable communication with desired QoS guarantees. Furthermore, each prototype may utilize different hardware and software flight stacks to realize the goal. The chosen hardware and software platforms may be open-source or proprietary in nature. Hence, the existing efforts tend to be very specific to the goal being pursued, thereby providing unique characteristics or behaviours to the prototype being developed. There are no formal development guidelines available so far in order to harmonize available features for these prototyping efforts.

An ideal view of cellular-connected UAV prototype is still missing. This ideal prototype can be thought of possessing a non-trivial list of mandatory features and should be adaptable to varying needs of the mission. Our current work attempts to foresight such an ideal prototype and enumerates the list of encompassing features. Table~\ref{relprotowork} illustrates a feature-oriented comparison of existing testbed works in literature with the desirable set of features from an ideal prototype point of view. Note that, this list of features is not exhaustive, rather provides a use-case driven analogy to consolidate the basic set of mandatory features. New features may arise in future with evolution of emerging use cases for cellular-connected UAVs. 

In this subsection, we aim at investigating the existing efforts to design and develop working prototypes for realizing some UAV operations over LTE/4G/5G/B5G cellular network infrastructure along with their implemented features. They are presented as follows.

\begin{figure}[b]
\centering
\includegraphics[width=1\linewidth]{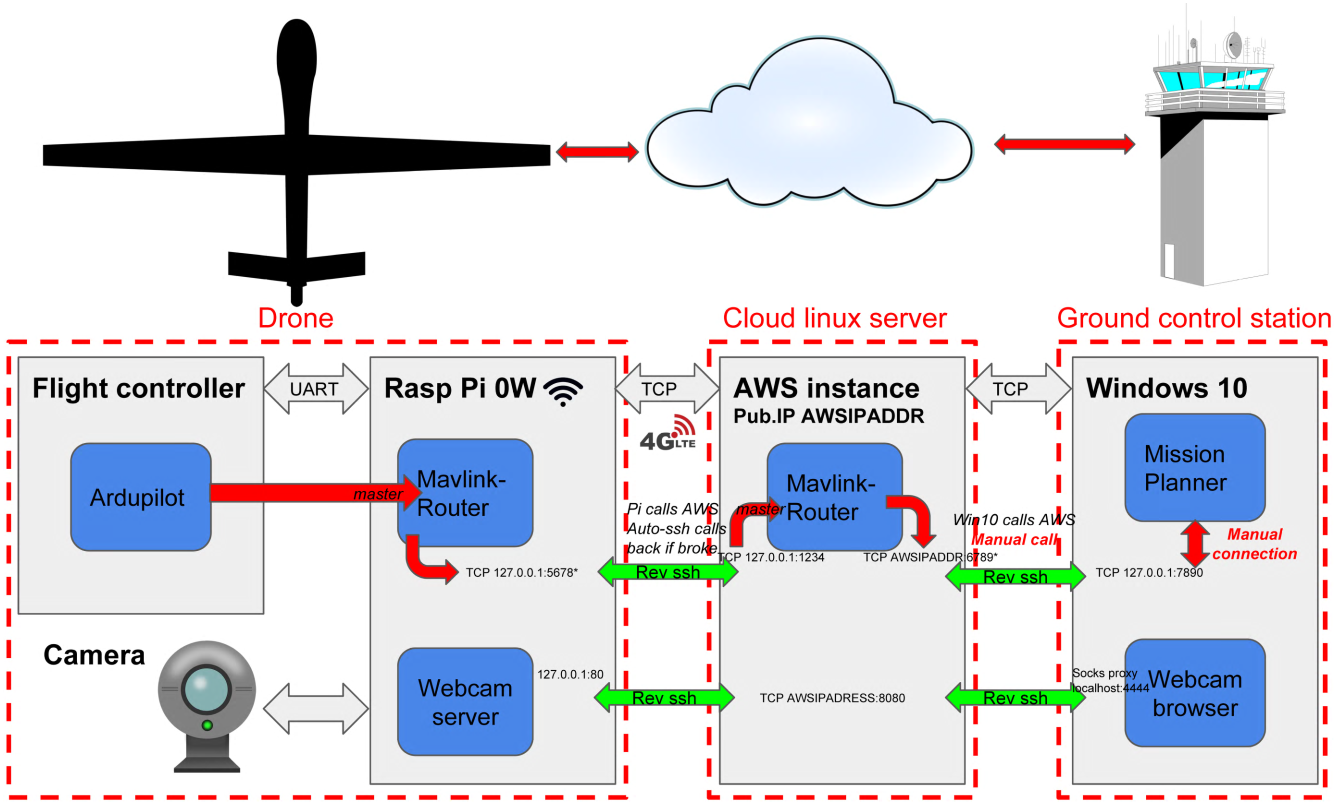}
\caption{Prototype design and configurations in~\cite{burke2019safe}}
\label{proto1}
\end{figure}

\begin{table}[width=0.9\linewidth,cols=2,pos=b]
  \caption{List of avionics components used in~\cite{burke2019safe}}
  \begin{tabular*}{\tblwidth}{@{} LL@{} }
   \toprule
    Component & Model\\
   \midrule
      Flight Controller & Omnibus F4 Pro \\  
	GPS & BN-220    \\
	Radio Rx & TBS Nano \\  
	Camera \& Video Tx & TX05    \\
	Computer & Raspberry Pi Zero W \\
	4G Modem & Verizon USB730L \\
	4G Antenna & TS9    \\
   \bottomrule
  \end{tabular*}
  \label{hwmodel}
\end{table}

\begin{figure}[b]
\centering
\includegraphics[width=0.9\linewidth]{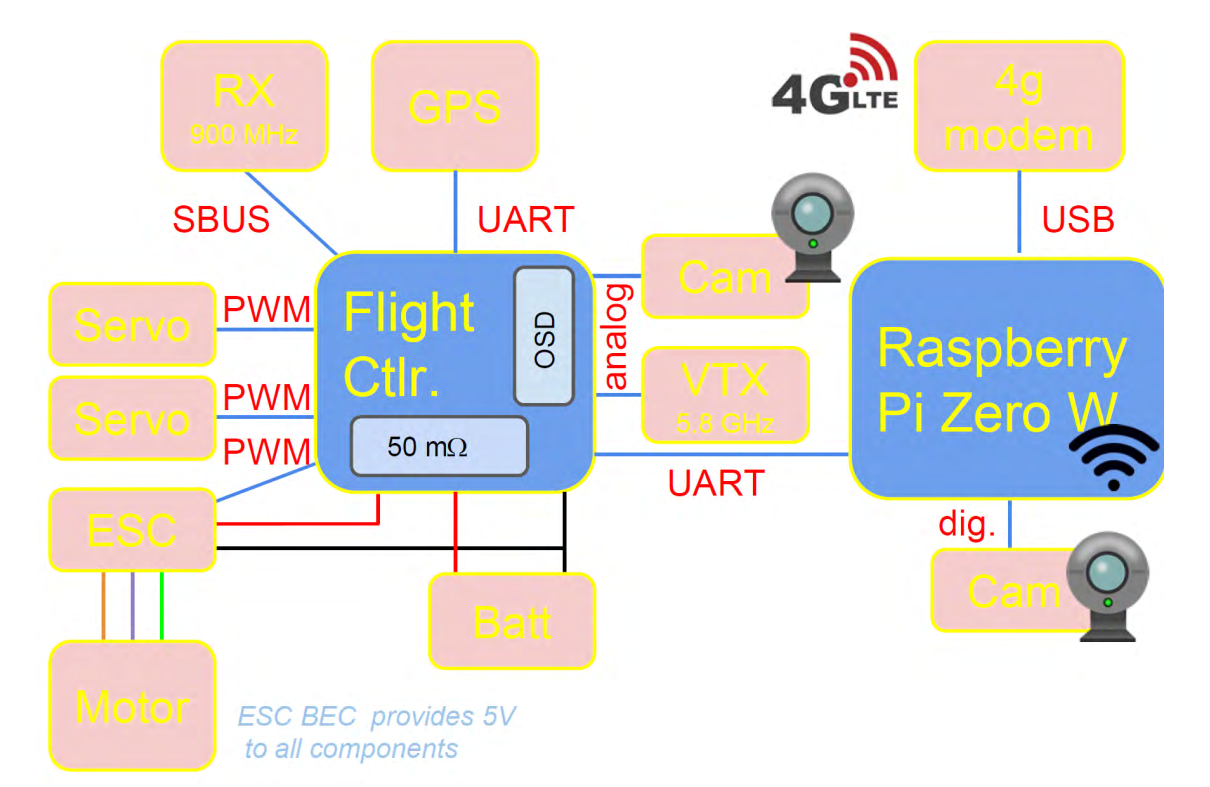}
\caption{Schematic of the avionics components in~\cite{burke2019safe}}
\label{hwschem}
\end{figure}

\begin{table*}[width=1\textwidth,cols=6,pos=h]
  \caption{A feature-oriented comparison of prototypes of cellular-connected UAVs from viewpoint of idealistic baseline}
  \begin{tabular*}{\tblwidth}{@{} CCCCCC@{} }
   \toprule
    \parbox{2cm}{\centering References \contour{black}{$\rightarrow$}\\Features \contour{black}{$\downarrow$}} & Short Description & ~\cite{burke2019safe} & ~\cite{sundqvist2015cellular} & ~\cite{solidakis2017arduino} & ~\cite{brodcnevs2018development}\\
   \midrule
\parbox{2cm}{\centering Cellular Network} & \parbox{8cm}{\vspace{.25\baselineskip}Cellular network generation type to which the prototype is connected and tested\vspace{.25\baselineskip}} &  \parbox{1cm}{\centering 4G\\LTE} & \parbox{1cm}{\centering 4G\\LTE} & \parbox{1cm}{\centering GSM/\\GPRS} & \parbox{1cm}{\centering 3G/4G\\LTE} \\
\midrule
\parbox{2cm}{\centering Open-source} &  \parbox{8cm}{\vspace{.25\baselineskip}Constituent hardware and software components of the prototype being developed\vspace{.25\baselineskip}} & \cmark & \cmark & \cmark &  \cmark \\
\midrule
\parbox{2cm}{\centering Autonomous} & \parbox{8cm}{\vspace{.25\baselineskip}Whether the UAV can fly autonomously without human intervention (self-flying nature)\vspace{.25\baselineskip}} &  \cmark &  \xmark &  \xmark & \xmark  \\  
\midrule
\parbox{2cm}{\centering Fail-safe} &  \parbox{8cm}{\vspace{.25\baselineskip}Ability to be resistant against lost link and returning to home location after UAV control is interrupted\vspace{.25\baselineskip}} & \cmark &  \xmark & \xmark  & \xmark   \\
\midrule
\parbox{2cm}{\centering Encrypted\\Communication} & \parbox{8cm}{\vspace{.25\baselineskip}Use of encryption mechanism to secure the message exchanges from potential attackers\vspace{.25\baselineskip}} & \cmark &  \xmark &  \xmark &  \xmark \\
\midrule
\parbox{2cm}{\centering BVLoS\\Capable} &  \parbox{8cm}{\vspace{.25\baselineskip}Being able to command and control the UAV, even not in the direct view of the remote pilot\vspace{.25\baselineskip}} &  \cmark & \cmark &  \xmark & \xmark \\
\midrule
\parbox{2cm}{\centering QoS-Aware} &  \parbox{8cm}{\vspace{.25\baselineskip}UAV successfully fulfils the application demands with respect to quality metric such as packet loss, latency, rate and jitter\vspace{.25\baselineskip}} &  \xmark &  \cmark & \xmark  & \xmark \\
\midrule
\parbox{2cm}{\centering Internet\\Connectivity} & \parbox{8cm}{\vspace{.25\baselineskip}Being able to control and steer UAV from persistent Internet connection\vspace{.25\baselineskip}} &  \cmark &  \xmark & \cmark &  \cmark \\
\midrule
\parbox{2cm}{\centering Ground\\Control} &  \parbox{8cm}{\vspace{.25\baselineskip}UAV being remotely controlled by ground control station for command and control, or payload communication\vspace{.25\baselineskip}} & \cmark & \cmark &  \cmark & \cmark  \\
\midrule
\parbox{2cm}{\centering Light-weight} &  \parbox{8cm}{\vspace{.25\baselineskip}Light-weight of UAV to enhance the prototype performance\vspace{.25\baselineskip}} & \cmark & \xmark & \xmark & \xmark  \\
\midrule
\parbox{2cm}{\centering Terrain\\Following} &  \parbox{8cm}{\vspace{.25\baselineskip}UAV maintains a fixed altitude and follows the terrain that is useful in unknown terrains like mountains\vspace{.25\baselineskip}} &  \cmark & \xmark & \xmark  &  \xmark \\
\midrule
\parbox{2cm}{\centering \vspace{.25\baselineskip}Flight\\Longevity\\($\sim$1 hour)\vspace{.25\baselineskip}} & \parbox{8cm}{\vspace{.25\baselineskip}Higher flight time of UAV indicating energy efficiency and negligible interruption during missions\vspace{.25\baselineskip}} & \cmark & \xmark &  \xmark & \xmark  \\
\midrule
\parbox{2cm}{\centering Endurance} & \parbox{8cm}{\vspace{.25\baselineskip}Robustness and integrity of UAV in extreme environment\vspace{.25\baselineskip}}  &  \cmark & \xmark & \xmark & \xmark \\
\midrule
\parbox{2cm}{\centering Energy\\Efficient} & \parbox{8cm}{\vspace{.25\baselineskip}Consumption of very less power to maintain persistent flight operation to accomplish the mission\vspace{.25\baselineskip}} & \cmark & \xmark & \xmark & \xmark \\
\midrule
\parbox{2cm}{\centering Network\\Virtualization} &  \parbox{8cm}{\vspace{.25\baselineskip}Ability of the prototype to be hardware platform independent and softwarization of UAV network functions\vspace{.25\baselineskip}} & \xmark & \xmark & \xmark & \xmark \\
\midrule
\parbox{2cm}{\centering Adaptable} &  \parbox{8cm}{\vspace{.25\baselineskip}Being responsive to current situation and the ability to reconfigure the UAV as per changing requirement and mission in minimum time\vspace{.25\baselineskip}} & \xmark & \xmark & \xmark & \xmark \\
\midrule
\parbox{2cm}{\centering AI/ML-Powered} &  \parbox{8cm}{\vspace{.25\baselineskip}Ability to leverage efficient AI or ML based approaches to self-learn and apply the learnt knowledge to improvise the mission performance over time\vspace{.25\baselineskip}} & \xmark & \xmark & \xmark & \xmark \\
\midrule
\parbox{2cm}{\centering Swarm\\Cooperation} &  \parbox{8cm}{\vspace{.25\baselineskip}Ability to properly coordinate information with other cellular-connected UAVs in a multi-UAV deployment scenario\vspace{.25\baselineskip}} &  \xmark & \xmark & \xmark & \xmark \\
   \bottomrule
  \end{tabular*}
  \label{relprotowork}
\end{table*}

An open-source 4G connected and controlled self-flying UAV is demonstrated in~\cite{burke2019safe}, defining a new, light-weight, secure and open-source class of cellular-connected UAV. This work utilizes open-source hardware and software stack to design and develop fully autonomous and fail-safe flight behaviour. This work provides a comprehensive and detailed discussion on the possible hardware and software options for flight controllers, radio receivers, sensors, microcontrollers and 4G cellular modems. Fig.~\ref{proto1} summarizes the hardware and software components used in the prototype development. The detailed hardware avionics schematics and equipment models are highlighted in Fig.~\ref{hwschem} and Table~\ref{hwmodel}, respectively. The performance of the prototype is tested for endurance, terrain alignment, autonomous flying behaviour, wind speed and real-time video quality. The important accomplishments of this work are summarized as follows.
\begin{itemize}
\item The entire prototype setup is done by open-source hardware and software components with Commercial off-the-shelf (COTS) components. 
\item The UAV shows longest demonstrated flight time i.e., over one hour. 
\item This work provides clear, concise and step-to-step guidelines for entire prototype design and development along with the programming of individual pieces. This also includes an online manual (wiki) and supplementary information.
\item The prototype shows self-healing internet architecture and utilizes the fail-safe protocols for the lost links in communication.
\item The UAV cellular to ground control is secure by encryption and can pass through several firewalls. 
\item As compared to other UAV industry verticals, the significant advantages are light weight (UAV weighs nearly 300 grams) and longer flight time (> 1 hour).
\end{itemize}

A working prototype of LTE controlled drone was demonstrated in~\cite{sundqvist2015cellular} proving the control of UAV via LTE connection and then tested as a 3D measurement platform. 
The goal of this prototype development was to investigate and evaluate LTE as a potential candidate of communication infrastructure for controlling a UAV. The experimental goals are to provide answer to below mentioned questions. 
\begin{itemize}
\item whether existing LTE network infrastructure is an efficient means of controlling UAVs?
\item whether the LTE connection is good enough in terms of providing low latency, jitter and bit error rate?
\item whether the bit rate is sufficient to perform the use case of live video streaming in BVLoS range?
\end{itemize}
The prototype is tested with respect to above mentioned goals and found that LTE is an efficient technology for UAV operations in BVLoS range satisfying the required the bit rate, latency and jitter. 
However, this prototype has several shortcomings and may not be considered as a full-fledged cellular-connected UAV testbed. Many features are either missing or not considered to keep the prototype simple in this development, thereby leaving enough scope for further enhancements. Some of the important features worth highlighting which are lacking in the prototype are listed below.
\begin{itemize}
\item The design did not consider cellular network coverage holes and discontinuity problem which may lead to failure of UAV operation. Flight fail safe mechanism is lacking.
\item The UAV mission specific investigation with respect to trajectory, interference from neighbouring base stations, handover criteria are missing from the design.
\item The QoS delivered by the UAV application must take into account diverse real-world use cases in presence of obstacles and variation of signal strength. Such study is missing.
\item It did not consider the factors and performance penalties when UAV coexist with other ground UEs.
\end{itemize}

\begin{figure}[t]
\centering
\includegraphics[width=0.8\linewidth]{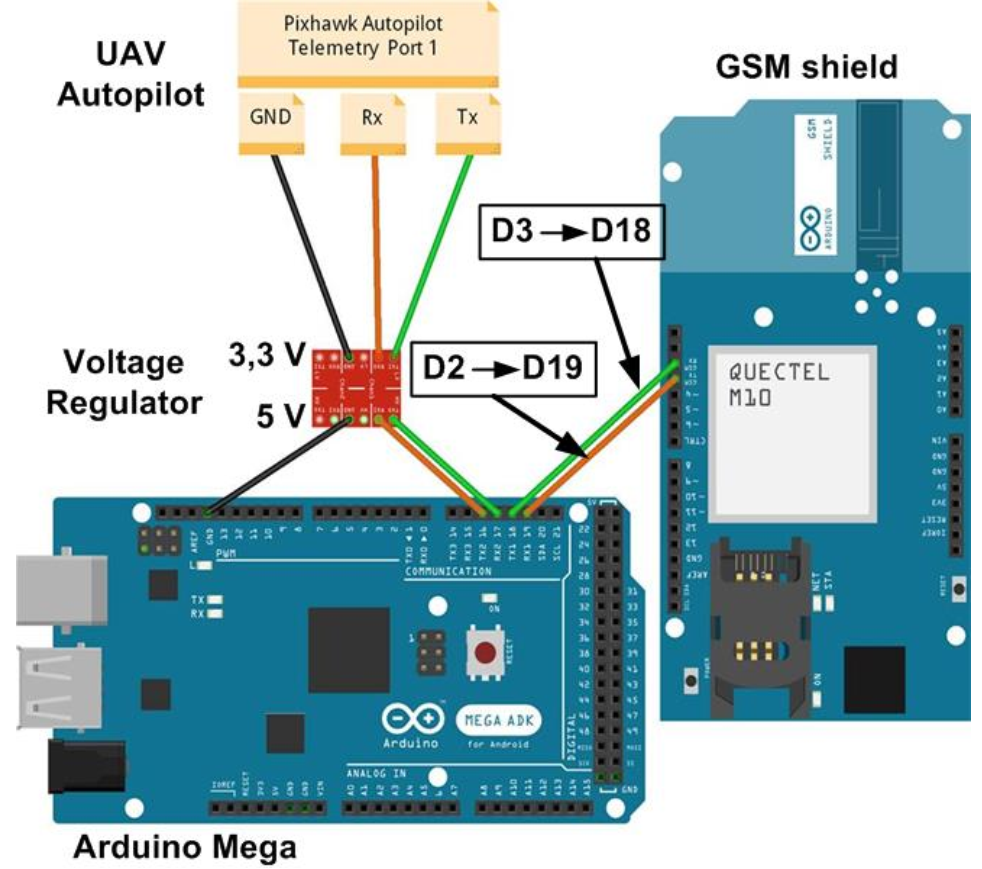}
\caption{High level shematics of the prototype setup in~\cite{solidakis2017arduino}}
\label{proto5}
\end{figure}

The work presented in~\cite{solidakis2017arduino} proposes an arduino-based low-cost, flexible control subsystem for controlling UAVs and ubiquitous UAV mission management by GSM/GPRS cellular networks. The ground control station transmits control signals to UAV present in LoS or beyond LoS over GSM or GPRS cellular network, through which, it is shown that is possible to connect to Internet, send/receive text messages or voice calls utilizing a GSM antenna and a SIM card. The experimental setup includes the following components: (i) UAV is an IRIS+ quadcoptor by 3DRobotics, (ii) Pixhawk autopilot, (iii) Arduino Mega ADK Rev. 3 microcontroller board, (iv) GSM/GPRS module by Arduino GSM shield with Quectel M10 modem, (v) Mission Planner, an open source software for ground control station software. The field tests are conducted by sending basic control commands from smartphone or laptop to UAV and they are successfully executed by the UAV. The subsystem initialization time is high, but occurs only once when the subsystem is powered ON. Fig.~\ref{proto5} shows the high level system schematics of the working prototype. Following are the key observations drawn from above experiment. 
\begin{itemize}
\item Communication via GPRS using a Mission Planner software has faster response time.
\item The Internet connectivity of GRPS is very fragile which make the GSM text message mode to be an efficient way for command and control message exchange.
\end{itemize}

A flexible open-source long-range communication solution for UAV telemetry based on cellular data transfer service is presented in~\cite{brodcnevs2018development} which is implemented on Raspberry Pi 3 model B (also known as rpi3) and Gentoo Linux control. The UAV is equipped with a Huawei 3372h dongle to get the cellular data services. \\


\subsection{Field Trials} 
In this subsection, various efforts on field trials and measurement campaigns of cellular-connected UAVs are discussed. Authors in~\cite{lin2019mobile} conducted a field measurement in a commercial LTE network for cellular-connected UAV operation. An LTE smartphone mounted on a consumer grade DJI Phantom 4Pro radio controlled quadcopter is used to gather the UAV flight results. The smartphone has TEMS Pocket 16.3 installed for wireless measurement and analysis. The field trial results include distribution and measurement of signal quality metric such as Reference Signal Received Power (RSRP), Reference Signal Received Quality (RSRQ), Signal to Interference and Noise Ratio (SINR) 
in the serving cell and neighbouring cells with respect to UAV movement. The results show the feasibility of UAV operations in commercial LTE network and also highlight the implementation challenges for dynamic radio environment.  The simulations are also conducted to supplement the field trial results in terms of network performance involving a higher number of cellular-connected UAVs. 
Following observations are drawn from the experiments.
\begin{itemize}
\item The aerial propagation conditions are close to free space propagation and hence, the aerial UEs experience stronger RSRP than the ground UEs.
\item The RSRQ and SINR at higher altitude is lower than the corresponding ground UE because of the strong downlink interference from neighbour non-serving cells to the aerial UE.
\item The uplink throughput for aerial UE is observed to be better than ground UE due to free space propagation condition. Note that, this uplink performance also depends upon many other factors like scheduling mechanism and network load.
\item In the downlink command and control traffic, higher altitude results in lower spectral efficiency due to increased interference and higher physical resource block (PRB) utilization.
\item Considering the mobility of aerial UEs at higher altitudes and LoS propagation conditions, a greater number of radio link failures (RLF) occur due to poor SINR and large interference. Also, they may connect to far-away cells instead of the closest cell.
\end{itemize}

In~\cite{raheeb2020uav}, the authors demonstrated an experimental platform where UAVs are connected to a commercial 5G NR base station for radio link measurements. The 5G BS is developed by Magenta Telekom in Austria and operates between 3.7 and 3.8 GHz frequency, using 100 MHz band. It has $64\times64$ massive MIMO setup with beam forming capabilities. An Asctec Pelican quadcopter is flown near this BS and the test measurements are performed by a Cellular Drone Measurement Tool (CDMT). This UAV carries a non-standalone Wistron NeWeb mobile test platform based on Qualcomm Snapdragon X50 5G modem. It supports sub-6 GHz 5G NR using $4\times4$ MIMO and 256-QAM. The goal of the study is to investigate the communication behaviour and performance characterization of flying UAV when connected to a commercially operated 5G base station. The communication aspects for 5G-connected UAV measured in this test are 5G connectivity, RSRP, SNR, throughput and number of handovers. The UAV flight includes both vertical lift-off and horizontal trajectory. Following observations are drawn from above testbed driven study of 5G connected UAV. 
\begin{itemize}
\item The UAV connectivity to 5G cannot be always guaranteed and fall back to 4G network. This situation is even worse at higher altitudes with more handovers towards 4G network.
\item The UAV is able to receive enough data rate (several hundred Mbps) from 5G NR based deployment, which is adequate for many applications and use cases.
\item The handovers to 4G network could be reduced by deploying a larger number of 5G NR base stations, and downlink rate would be improved. However, the experiment did not yield much benefit in the uplink as compared to 4G. The authors assume that uplink rate analysis needs further investigation.
\end{itemize}

Qualcomm also tested the UAV operation in commercial LTE networks in September 2016 and produced a trial report in May 2017 on LTE unmanned aircraft system~\cite{qualcomm2017unmanned}. The focus of this test was to understand the operation of low altitude UAV platforms being supported by terrestrial cellular networks. The overall test encompasses both field trials and simulations. The field trials aim to capture datasets by performing hundreds of flights and then complemented by extensive system level simulations to understand the performance of UAV operation. The flights and measurements were performed by custom designed 390QC quadrotor drone.  Note that, these results are collected in a suburban/residential zone which was having good cellular coverage, hence, cannot be generalized for other zones like urban or rural areas. Moreover, the performance results are approximate in nature rather than accurate. The key results obtained from the trail report are summarized as follows.
\begin{itemize}
\item The aerial UEs experience higher received signal strength than ground UEs despite of the down-tilted BS antennas. This is because of the better free space propagation condition at higher altitude.
\item The SINR in the downlink for aerial UEs is lower as compared to ground UEs due to the interference experience from neighbour cell.
\item The UE transmit power is more for ground UEs than aerial UEs in the uplink, because good free space propagation condition at higher altitude enhances the interference energy from neighbour cell. The field results depicted that aerial UEs experience nearly three times more interference than ground UEs in 700 MHz band.
\item Handover performance in terms of lower handover frequency and success rate of handovers is superior for aerial UE than ground UE due to signal stability at high altitude.
\item The optimization in the power control scheme are applied by simulation and was shown to eliminate the excess uplink interference. 
\end{itemize}

The work in~\cite{amorim2017pathloss} presents the field trial done at a small airport in vicinity of Odense, Denmark and results were collected in an LTE network operating in 800 MHz and the UAV altitude is maintained between 20 to 100 meters. The cellular network data was collected by a Samsung Galaxy S5 smartphone which was placed inside the flying UAV cavity. It was equipped with Qualipoc software for reporting the radio measurements. The UE was programmed to use a 20MHz carrier with centre frequency at 810 MHz. The measurement field had no obstruction between UAV and BS and was an open area to prevent significant signal attenuation and reflected paths. For every second interval, the software radio reports include RSRP and RSRQ. The goal of the experiment was to understand the effects of altitude on the radio performance of an aerial user. It is observed that the SINR degrades when the UAV flies high. The steepest degradation is seen in height variation between 20-40 metres indicating that the increase in interference is more prominent at lower altitudes, and smaller variations at higher altitudes due to improved gain. The study reveals that, expanded radio horizon at higher altitudes, high LoS probability and clearance of first Fresnel zones are key radio factors in modelling the path loss between aerial users and ground stations.

In~\cite{bertizzolo2020live}, the authors experimentally evaluated the terrestrial users’ performance in the presence of UAV as aerial users on LTE testbed. The performance measurements are conducted to measure the throughput degradation. The LTE network is considered to operate over 2300 MHz carrier frequency with 20 MHz operational bandwidth spanning an area of 160000 square feet. The setup includes 2 eNodeBs, 4 LTE cells each having $2 \times 2$ MIMO capability. The downlink and uplink bitrates are kept as 150 Mbps and 50 Mbps, respectively. The UAV hovers at a height of 50 meters. To analyse the performance, both ground UE and aerial UE generate uplink traffic at full buffer capacity for one minute in each experiment run. It is observed that the ground UE suffers a throughput degradation up to 21.75 Mbps because of the inter-cell uplink interference by aerial users. The average reduction in throughput is nearly 52\% i.e., equivalent to 11 Mbps.

The work in~\cite{hayat2019experimental} presents the experimental evaluation of cellular-connected UAVs communication performance connected to an LTE-Advanced network running 3GPP Release 13. An Asctec Pelican quadrocoptor carrying a smartphone (Sony Xperia XZ2 H8216) flies in the coverage of an LTE-A network within the premises of University of Klagenfurt campus. The experiment is performed in open-field and obstacle-free areas ensuring LoS link with at least one BS. The UE supports LTE carrier aggregation and a $2 \times 2$ MIMO antenna setup is used. The base station has a transmit power of 20 Watt and 256 QAM and 64 QAM in downlink and uplink, respectively. The UE was able to report various LTE parameters such as RSRP, RSRQ, SINR, serving PCI, TCP uplink and downlink throughput, EARFCN etc. The UAV followed a straight path spanning 300 meters with a speed of 3 meters/second. The broad goal of this experimental study was to understand the impact of varying UAV altitude on achievable throughput and performance measurement of handovers by aerial user without any specific change in the network. The keys findings are as follows:
\begin{itemize}
\item The achievable throughput of UAV is sufficient enough to cater to many applications and use cases. At an altitude of 150 metres, the UAV’s average throughput is 20 Mbps and 40 Mbps in the downlink and uplink, respectively.
\item The number of handovers increase with increasing height of UAV. The reason for high handover frequency is the high RSRP and high interference values from neighbour base stations. 
\end{itemize}

In~\cite{amorim2017machine}, the authors used machine learning algorithms in order to identify the cellular-connected UAVs in the network based on LTE radio measurements. The measurement was conducted in a rural location of Northern Denmark where the airborne aerial UAV users are realized by mounting a QualiPoc android smartphone on commercial UAV attached to an 800 MHz LTE carrier. The height is maintained at 4 different levels and UAV is flown in 4 rectangular routes. This work claims the use of supervised learning algorithms for efficient detection of aerial users in the network solely based on LTE radio measurements with small number of training samples.

Authors in~\cite{amorim2018lte} focuses on aerial communication field trial, where a radio scanner is attached to construction lift and radio signal was measured with heights up to 40 metres in urban scenario. The measurement was carried out in three LTE carriers such as 800, 1800 and 2600 MHz in northern Denmark. The experimental study aims at providing propagation models of UAVs connected to cellular networks. The key findings from the trials are as follows:
\begin{itemize}
\item Increase in the received power from neighbour sources even in height of 40 meters that contributes to heavy interference for the aerial user.
\item The observed path loss approximated to free space path loss after a UAV height of 25 meters.
\end{itemize}

The authors in~\cite{marques2019experimental} have demonstrated the feasibility of UAV operation via commercial cellular network for high data connectivity in low altitude and BVLoS operations throughout different times of the day. 

\begin{table*}[width=1\textwidth,cols=4,pos=t]
  \caption{Comparison of existing works on field trials and measurement campaigns}
  \begin{tabular*}{\tblwidth}{@{} CCCC@{} }
   \toprule
    Reference Work & Cellular Network & Trial Environment & Performance Metric \\
   \midrule
   ~\cite{lin2019mobile} & 4G LTE & Suburban & \parbox{8cm}{\vspace{.25\baselineskip} RSRP, RSRQ, SINR, downlink latency, resource utilization  \vspace{.25\baselineskip}} \\ 
~\cite{raheeb2020uav} & 5G New Radio & Rural &\parbox{8cm}{\vspace{.25\baselineskip} RSRP, SNR, Throughput, 5G connectivity  \vspace{.25\baselineskip}} \\ 
~\cite{qualcomm2017unmanned}  & 4G LTE & Mixed Suburban &\parbox{8cm}{\vspace{.25\baselineskip} Cellular connectivity for low altitude UAVs  \vspace{.25\baselineskip}} \\ 
~\cite{amorim2017pathloss} & 4G LTE & Rural & \parbox{8cm}{\vspace{.25\baselineskip} RSRP, RSRQ, SINR, Effect of altitude on UAV   \vspace{.25\baselineskip}} \\ 
~\cite{bertizzolo2020live} & 4G LTE & Rural, Suburban, and Urban & \parbox{8cm}{\vspace{.25\baselineskip} Throughput degradation, Interference, Uplink signal power  \vspace{.25\baselineskip}} \\ 
~\cite{hayat2019experimental}  & 4G LTE-A & Unknown & \parbox{8cm}{\vspace{.25\baselineskip} RSRP, RSRQ, SINR, PCI, UL/DL throughput, EARFCN  \vspace{.25\baselineskip}} \\ 
~\cite{amorim2017machine} & 4G LTE & Rural & \parbox{8cm}{\vspace{.25\baselineskip} Cellular-connected UAV identification  \vspace{.25\baselineskip}} \\ 
~\cite{amorim2018lte} & 4G LTE & Urban & \parbox{8cm}{\vspace{.25\baselineskip} Channel propagation models  \vspace{.25\baselineskip}} \\ 
~\cite{marques2019experimental}  & 4G LTE & Rural & \parbox{8cm}{\vspace{.25\baselineskip} RSSI, RSRP, RSRQ, uplink/downlink throughput  \vspace{.25\baselineskip}} \\ 
   \bottomrule
  \end{tabular*}
  \label{relmeaswork}
\end{table*}

Table~\ref{relmeaswork} presents a comparative analysis of different existing works in literature with respect to field trials and measurement campaigns. Existing field trials vary greatly in several aspects, such as type of environment, deployment scenario, modelling platform, goal of experiment, performance metric, etc.

\subsection{Summary of Lessons Learnt}
The important lessons learnt from this section are listed as follows:
\begin{itemize}
\item The solutions proposed to the technical synergistic challenges of 5G/B5G systems with UAV technology must be validated and tested for correctness. Hence, the key design considerations of cellular-connected UAV necessitate sound measurement campaigns, field trials, simulations and working prototypes to study the behaviour with real-world scenarios. The experimental testbeds and field measurements have significant practical relevance, because these are very conducive for realistic evaluation of the system under study.
\item Most of the existing literature relies on software simulations to evaluate the technical aspects of proposed solutions. Few works have shown to conduct field measurement campaigns to observe and study the behaviour of cellular-connected UAV such as handovers, cell association, signal strength reduction, radio link status, interference mitigation etc. Very few works focus on design and development of real-world working prototypes of cellular-connected UAV, because such efforts are time-taking and highly complex. Prototyping works are still in its infancy stage and hence, call for more contributions in this regard.
\end{itemize}

\section{Standardization \& Socio-economic Concerns} \label{challenges}
Cellular-connected UAVs can pose serious risks in terms of socio-economic operational capabilities. Therefore, utmost care must be taken by the policy makers and legislation in order to integrate UAVs into national and international aviation systems. To this end, in this section, we outline the perspectives of standardization, regulatory activities, market and social challenges, which the UAV service providers and cellular operators must take into consideration before successful roll-out of use cases pertaining to cellular-connected UAV applications.

\subsection{Standardization} 
Third Generation Partnership Project (3GPP) is a standardization body that governs the specifications for the technical platforms used by the cellular networks. The global partnership 3GPP develops standards to which almost all commercial cellular network providers and operators adhere to. In order to cater to the present and future needs of UAV communication, 3GPP aims to layout a unified platform for design and development of wireless innovations by gaining wider consensus from various contributors from industry and academia. The evolutions in the standards are published in the name of ``3GPP release". From the perspective of UAV operations over cellular networks, we are interested in 3GPP Release-15, Release-16 and Release-17. In Release-15, the study mainly concerns with the radio level aspects for supporting UAVs. In Release-16 and 17, the study is in the perspective of System and Application layer aspects. \\

\subsubsection{Release-15} 
3GPP led a study item (SI) in Release-15 to investigate various prospects of utilizing an LTE network for UAV communication. The key outcomes of this study are summarized in the technical report TR 36.777, which was approved in January, 2018. This study focuses on two broad goals: First, how the aerial users (cellular-connected UAVs) impact overall LTE performance in the presence of terrestrial (ground) users. Second, to investigate on whether an LTE network is able to provide good support low altitude UAVs? The comprehensive list of items studied in this release are: channel modelling between aerial UEs and ground BSs, uplink and downlink interference problems due to LoS channel propagation characteristics, identification of aerial UE for legitimate cellular usage and subscription information, mobility performance, and flight path signalling. The result of the SI shows that existing LTE networks are able to support UAV communication and there is no notable impact in coexistence of small number of aerial and ground users (low density or rural regions). However, increase in the number of aerial and ground users have adverse impact on uplink/downlink performance due to interference. To some extent, existing LTE standards are found useful to mitigate the interference situation. After the completion of aforementioned study, a work item (WI) is pursued and got approved for enhancement of LTE standards. They are as follows:
\begin{itemize}
	\item Introduction of new radio events and enhanced height dependent reporting for aerial UEs;
	\item Support of signalling in subscription based aerial user identification;
	\item Improvement of mobility and interference detection, uplink power control, airborne status and flight path plan.
\end{itemize}

\subsubsection{Release-16} 
\begin{figure}[t]
\centering
\includegraphics[width=1\linewidth]{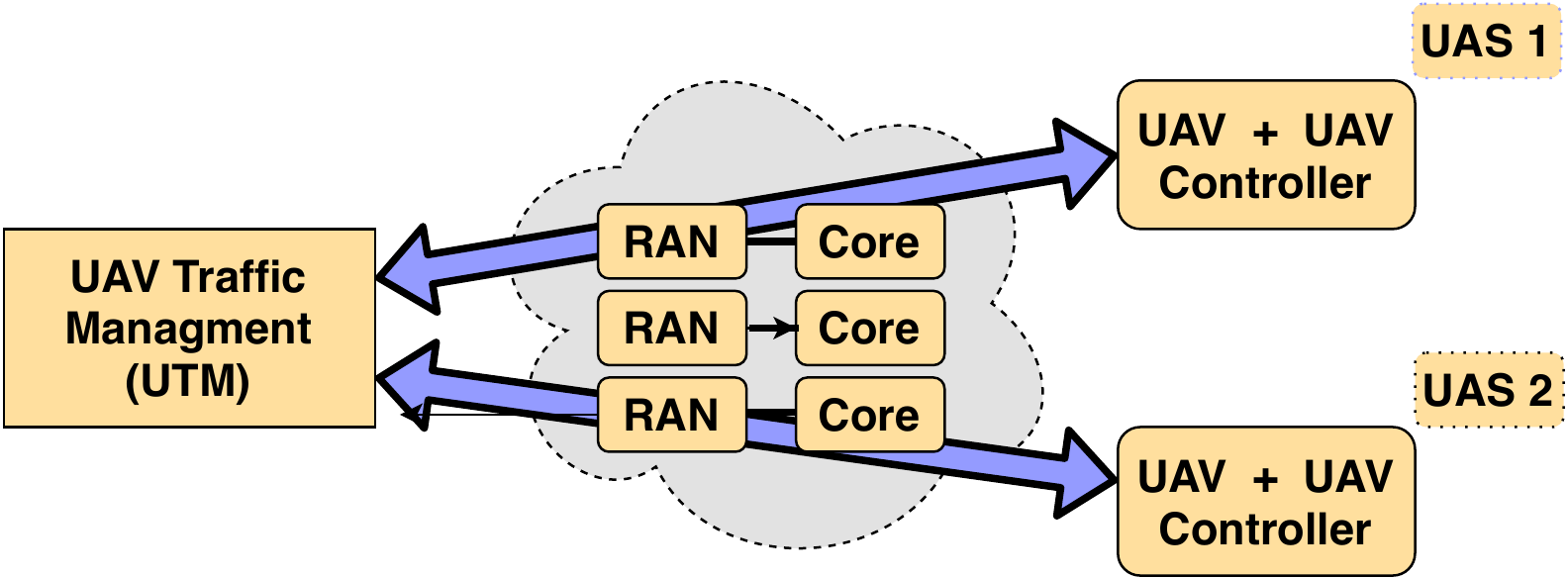}
\caption{High level architecture of 3GPP Release 16 work on remote identification of UAS}
\label{rel16}
\end{figure}

This release plan started in September, 2016 and the approval for stage-3 development was conducted on December, 2019. This release work is mainly on System and Application layer aspects. A study item on “Remote Identification of Unmanned Aerial Systems” is pursued on this release and it led to the approved report 22.829. This study aims at the identification of UAV over the command and control data via a 3GPP network exchanged between UAS and centralized UAV Traffic Management (UTM) component. A UAS comprises of UAV and UAV controller. Fig.~\ref{rel16} depicts above model. 3GPP standards must make provisions for UAS to send the application data traffic to UTM along with various radio network information, identification and tracking details for UAS. After this study, a WI was agreed by 3GPP to advance the work on service requirement for identification of UAV. \\

\subsubsection{Release-17} 
In this release, 3GPP proposes a number of study items. The idea is to come up with diverse scenarios and metrics to cater to wide variety of UAV applications and use cases. The study items are as follows:
\begin{itemize}
	\item 5G Enhancement for UAVs :- It includes several Key Performance Indicators (KPIs) relevant to UAV services. The KPIs are provided for command and control, and payload communication.
	\item Study on application layer support for Unmanned Aerial System :- This includes the UAS service requirements that may have impact on the UAS application layer. These requirements are in terms of general requirements, UE capability identification, location, security etc.
	\item Study on supporting Unmanned Aerial Systems Connectivity, Identification, and Tracking :- This study item deals with a mechanism that enables the UAS tracking and identification within 3GPP systems and UTM.
\end{itemize}

\subsection{Regulatory Concerns} 
Ubiquitous accessibility and rapid emergence of UAV technology mandate development of regulatory frameworks for harmonious operation of cellular-connected UAVs in the national and international airspace. Although, each country has a specific set of internal rules for UAV operation, few global bodies tend to harmonize their operation across international airspace. The regulatory framework mainly target around three key aspects~\cite{stocker2017review}:
\begin{itemize}
	\item To regulate and control the use of unmanned aircraft in the airspace to prevent danger to manned aircraft;
	\item To ensure proper operational limitations to the flight;
	\item To manage and control the administrative privileges such as pilot licenses, flight authorizations and data handling techniques.
\end{itemize}

\subsection{Market Concerns} 
The UAV ecosystem leveraging the emerging technologies such as IoT, AI, AR/VR are not much explored by the manufacturers and their usages are also researched by a handful of organizations. Real-time surveillance is one of the major use case that has been widely explored by the UAV industry for relaying live information to target audience. The ecosystem is still in its infancy to showcase diverse capabilities of cellular-connected UAVs. Additionally, the skills necessary for UAV industry to roll out interesting use case demand sufficient domain training to the equipment providers and technical users. It is key to eradicate the bottleneck in setting up the ecosystem. Extracting the right set of specifications and requirements from the users is needed to maximize the benefit of the use case and to generate large scale development of cellular-connected UAV applications.

\subsection{Social Concerns}
The UAV operation must be properly regulated to protect the privacy of business organizations as well as individuals. Advancement of drone technology with aerial surveillance and photography with high definition images and streaming can easily violate the privacy, even when it is unintentional. The existing regulations to protect privacy may not be sufficient due to rapid evolution of UAV technology and its increasing capability, thereby further legislation is needed to be formulated to protect privacy. 

Most of the UAV use cases deal with gathering a lot of vital data depending upon the application and processing them to extract useful information for taking decisions. During a mission, the onboard sensors collect personal or business data can be transmitted to a remote location or made live from the present location. The data collection capabilities may infringe data protection rules and abuse personal information without the knowledge of data subject. Additionally, if someone obtains the control of the UAV, the sensors or data processing circuitry could be tempered for data misuse. Hence, strict guidelines must be governed to protect the personal and business data. 

In the due course of flight or mission control, any sort of discontinuity in proper command and control poses serious safety risks. This may lead to collisions and causes harm to civilians and other UAVs in the vicinity. Collisions with manned aircraft can pose serious risks in terms of catastrophic consequence and loss of assets. In case of high density urban regions, collisions of UAVs with the ground terrain pose threats to human lives and assets. Hence, the challenges with respect to public safety must be taken into account and researched thoroughly.

\subsection{Summary of Lessons Learnt}
The important lessons learnt from this section are listed as follows:
\begin{itemize}
	\item Cellular-connected UAVs not only impose technical challenges, but also require solutions pertaining to privacy, security, licenses, public safety, administrative procedures governing them. Standardization bodies such as 3GPP have put together study items and working groups in order to harmonize the development efforts from industry, academia and independent research bodies. 
	\item Operation of UAVs over cellular spectrum requires strict regulations to operate in national and international airspace without causing trouble to other manned or unmanned aerial vehicles. There are rules applied to control UAV operation that varies with countries. However, a unified set of rules governing UAV operation in cellular spectrum is still far away. 
	\item The commercial production of cellular-connected UAVs must consider the true requirements and specifications to maximize the benefit of a use case.
	\item Care must be taken to safeguard the data collected by the sensors of UAV and ensuring that it does not infringe the privacy of unwanted individuals and organizations. It must be guarded against hackers and malicious intruders, whose intent is to control the UAVs for unauthorized activity, e.g., during aerial surveillance and photography.
\end{itemize}

\section{Future Outlooks} \label{futout}
In earlier sections, we have outlined integration challenges and highlighted candidate 5G/B5G innovations to address some of those challenges for cellular-connected UAV. Despite of several studies, there is still considerable areas of open problems that needs to be investigated. The current section aims to bring out such future opportunities for researchers and shed light on interesting open research topics.

\subsection{Accurate Channel Models}
UAVs are expected to be deployed in a wide variety of indoor and outdoor environments such as stadiums, urban, rural, sub-urban, industrial, over water, highways etc. All these environments require unique air-ground propagation conditions, measurement campaigns and channel models for cellular-connected UAVs. However, accurate modelling and characterizing each of these unique channels is a non-trivial task that remains largely unexplored. In addition, introduction of 5G-oriented technologies such as mmWave and massive MIMO systems bring additional factors and scope for improvements in design and development of effective channel models for cellular-connected UAVs. 

\subsection{Energy/Battery Power Limitations}
Onboard energy is a bottleneck in UAV. Recent developments to rechargeable battery cells and use of solar cells are some of the ways to extend the flying time of UAV. The UAVs require continuous power source to operate as they use a huge percentage of battery power in flying. Different actions such as transmission, reception, execution of software function, path planning optimizations consume the battery power. Most of the existing works on cellular-connected UAV do not factor this energy limitation during the study. There exists a lot of scope to investigate the performance of cellular-connected UAVs considering the limited energy constraints especially in areas of VNF deployment on UAVs for automated operation, trajectory optimization, learning-based methods, and longevity calculation of the mission in a use case. 

\subsection{Security and Privacy}
Due to open communication links, the cellular-connected UAVs are vulnerable to cyber-physical or malicious attacks to spoof the control signals. Such attempts pose tremendous threat to the UAV system in terms of loss/stealing of confidential information or failure of mission. The signal spoofing of control signal might have adverse effect on the UAV mission and making it difficult to bring it back online. Hence, in order to avoid such malicious modifications, a relevant open issue is to improvise the security and privacy aspect of UAV cellular communication that demands in-depth study of security issues spanning all layers of the protocol stack. 

\subsection{Cell Selection and Network Planning}
The UAVs are typically served by the side lobes of current base stations in 4G and 5G cellular networks and therefore may easily establish cell selection with far away base stations. Furthermore, the high mobility of UAVs increases the failure rates of radio link due to more frequent handovers. The design of antennas to supports flying UAV in high altitude and enhanced solutions for radio network planning are required by network operators and equipment vendors. 

\subsection{3D Mobility and Handovers}
Typically, ground UEs are mobile in 2D space and base station antennas are optimally designed for ground users. UAVs are typically served by the side lobes of current base station and their aerial patterns are different resulting in unique handover characteristics, different from ground UEs. The frequent handover pattern is largely dependent on blockage, mobility, altitude variations. There is a need for improved handover mechanisms that suit the characteristics of high mobility UAVs. Additionally, the cell selection schemes based on nearest or strongest RSRP may no longer be an appropriate method for cellular-connected UAVs.  

\subsection{Testing platforms}
Simulations and measurement campaigns are not sufficient to fully characterize the performance and working principles of cellular-connected UAVs. The proposed solution approaches to the integration challenges must be complemented by extensive field trials and real-world testbed-based evaluation. However, there are not enough working prototypes to study and evaluate the behaviour of cellular-connected UAVs from practical standpoint. Additionally, there are no experimental or open-source simulator platform available till date to assist in a wide range of functionality testing of UAVs. Such works need to be pursued in future to fill in the gap between theoretical proposal and practical evaluation. 

\subsection{AI and ML-based methods}
AI and ML-based approaches have been considered as powerful tools to solve many real-world wireless networking problems revolutionizing 5G/B5G networks. On this advent, the UAV cellular communication has opened up new possibility for autonomous UAV operation considering the security, performance and dynamic complex deployment scenarios. Numerous research issues exist in studying and evaluating AI/ML-empowered techniques to solve challenges of UAV-ground interference management, power control, multi-UAV cooperation etc. Moreover, Q-learning approaches are helpful for UAVs to tackle security issues by adaptively controls the UAV transmission as per the malicious type of attacks. 

\subsection{Computational Offloading}
Due to moderate computational capabilities and limited onboard energy of UAVs, MEC can be helpful for offloading computationally heavy tasks from cellular-connected UAV to edge nodes to improve endurance and life time of UAV. Some examples of such intense tasks are real-time face recognition in a crowd surveillance use case. For such use case, leveraging MEC capabilities along with UAV, the recognition task can be offloaded to complete the job in a timely manner. Additionally, to save the information from eavesdropper, proper security mechanisms needs to be integrated to the MEC-UAV platform for optimum performance. These research areas are largely unexplored so far and numerous scopes exists for design and development of UAV-MEC frameworks for cellular-connected UAVs considering its diverse use case.

\section{Conclusions} \label{cnwf}
In this work, we provided a comprehensive study on the Cellular-assisted UAV communication paradigm (Cellular-connected UAV) where UAV is integrated to existing 5G/B5G cellular systems as a new aerial UE. The detailed taxonomy of various application domains with emerging use cases as well as the technical synergistic challenges of UAV integration with cellular network are discussed first. Then, we focus on the promising network architectures and physical layer improvements in 5G/B5G systems considering the hardware and software design challenges of Cellular-connected UAVs. The key innovative 5G technologies are elaborated enabling the seamless integration and support of UAV communication over cellular spectrum. In order to characterize the design performance benefits and study the realistic deployment issues, we also highlighted the efforts to develop working prototypes as well as the field trials and simulations. The progress on standardization activities by 3GPP, national and international regulations and concerns pertaining to socio-economic barriers are also discussed which must be accounted before successful adoption this new technology. We believe this work will be a very useful and motivating resource for researchers working on cellular-connected UAVs in order to unlock a holistic view and to exploit its full potential.


\bibliography{bib_file}

\end{document}